%% file: kththesis.tex
\documentclass{kthreport}

\usepackage{blindtext} 

\usepackage{csquotes} 
\usepackage[style=numeric,sorting=none,backend=biber]{biblatex}
\usepackage{hyperref}
\usepackage{color, colortbl}
\usepackage{tikz}
\usepackage{subcaption}
\usepackage{listings}
\usepackage{acro}
\acsetup{first-style=short}

\definecolor{mGreen}{rgb}{0,0.6,0}
\definecolor{mGray}{rgb}{0.5,0.5,0.5}
\definecolor{mPurple}{rgb}{0.58,0,0.82}
\definecolor{backgroundColour}{rgb}{0.95,0.95,0.92}
\lstdefinestyle{CStyle}{
    backgroundcolor=\color{backgroundColour},   
    commentstyle=\color{mGreen},
    keywordstyle=\color{magenta},
    numberstyle=\tiny\color{mGray},
    stringstyle=\color{mPurple},
    basicstyle=\footnotesize,
    breakatwhitespace=false,         
    breaklines=true,                 
    captionpos=b,                    
    keepspaces=true,                 
    numbers=left,                    
    numbersep=5pt,                  
    showspaces=false,                
    showstringspaces=false,
    showtabs=false,                  
    tabsize=2,
    language=C
}

\usetikzlibrary{shapes.geometric, arrows}
\tikzstyle{process} = [rectangle, minimum width=1.7cm, minimum height=0.75cm, text centered, text width=3cm, draw=black]
\tikzstyle{input} = [rectangle, minimum width=1.7cm, minimum height=0.75cm, text centered, text width=3cm, draw=black]
\tikzstyle{arrow} = [thick,->,>=stealth]
\tikzstyle{process1} = [rectangle, minimum width=1.7cm, minimum height=0.75cm, text centered, text width=1.5cm, draw=black]
\tikzstyle{process2} = [rectangle, minimum width=1.0cm, minimum height=0.75cm, text centered, text width=1.0cm, draw=black]
\tikzstyle{input1} = [rectangle, minimum width=1.7cm, minimum height=0.75cm, text centered, text width=1.5cm, draw=black]
\tikzstyle{input2} = [rectangle, minimum width=1.1cm, minimum height=0.75cm, text centered, text width=1.1cm, draw=black]
\tikzstyle{process_lossless} = [rectangle, minimum width=1.8cm, minimum height=0.75cm, text centered, text width=3cm, draw=black]
\tikzstyle{input_lossless} = [rectangle, minimum width=2.1cm, minimum height=0.75cm, text centered, text width=2.0cm, draw=black]

\title{Multi-resolution encoding and optimization for next generation video compression}
\alttitle{Kodning och optimering med flera upplösningar för nästa generations videokomprimering}
\author{Vignesh V Menon}
\email{vvba@kth.se}
\supervisor{Jonatan Samuelsson}
\examiner{Markus Flierl}
\hostcompany{Divideon AB} 
\programme{MSc Information and Network Engineering}
\school{School of Electrical Engineering and Computer Science}
\date{October 16, 2020}

\kthcover{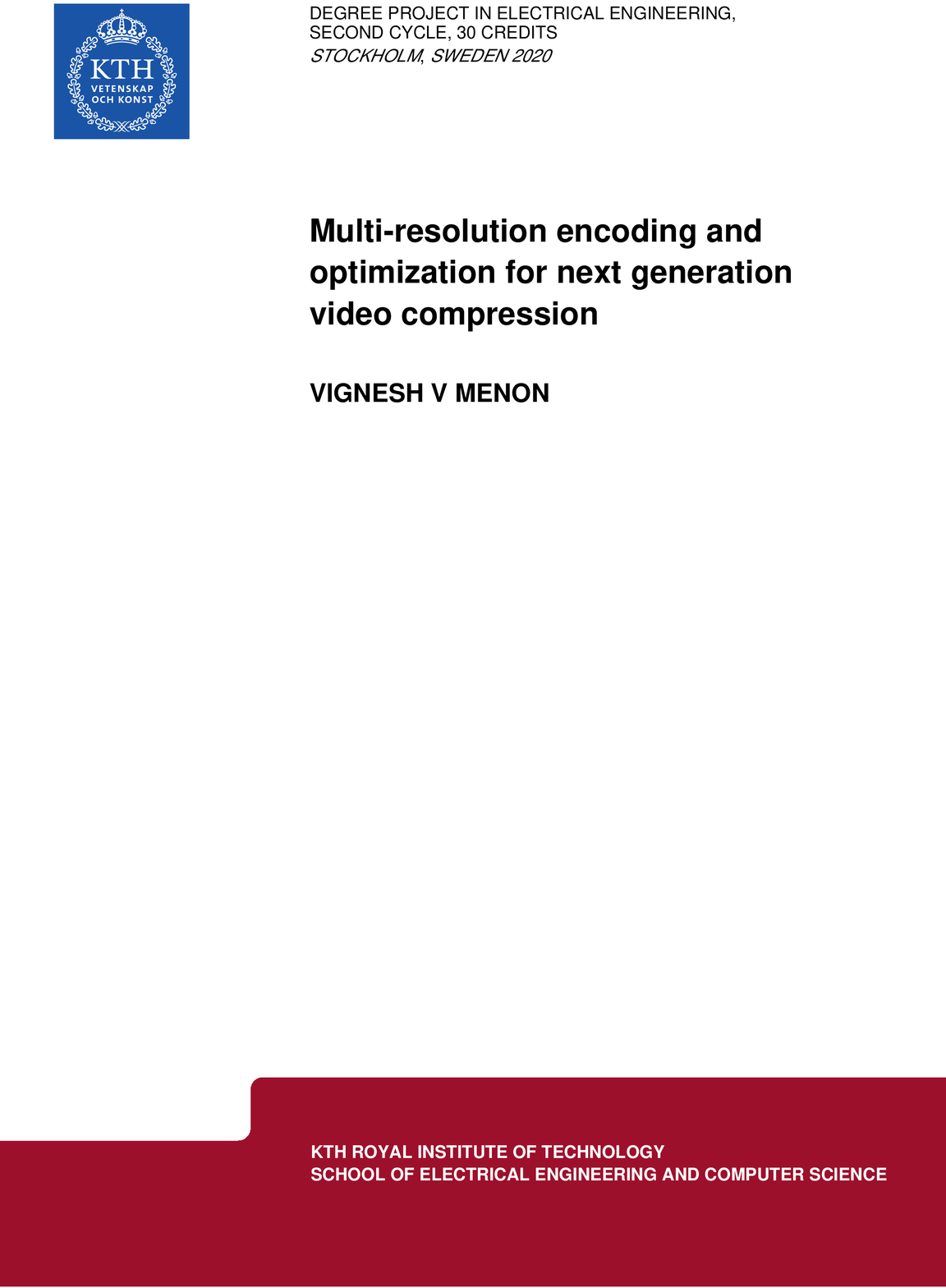}

\DeclareAcronym{hevc}{
  short = HEVC ,
  long  = High Efficiency Video Coding ,
  tag = abbrev
}

\DeclareAcronym{abr}{
  short = ABR ,
  long  = Adaptive bitrate ,
  tag = abbrev
}

\DeclareAcronym{http}{
  short = HTTP ,
  long  = Hypertext Transfer Protocol ,
  tag = abbrev
}

\DeclareAcronym{cvbr}{
  short = cVBR ,
  long  = Capped Variable Bitrate ,
  tag = abbrev
}

\DeclareAcronym{cqp}{
  short = CQP ,
  long  = Constant Quantization Parameter ,
  tag = abbrev
}

\DeclareAcronym{cpu}{
  short = CPU ,
  long  = Central Processing Unit,
  tag = abbrev
}

\DeclareAcronym{uhd}{
  short = UHD ,
  long  = Ultra High Definition,
  tag = abbrev
}

\DeclareAcronym{hdr}{
  short = HDR ,
  long  = High Dynamic Range,
  tag = abbrev
}

\DeclareAcronym{vod}{
  short = VOD ,
  long  = Video On Demand,
  tag = abbrev
}

\DeclareAcronym{dvd}{
  short = DVD ,
  long  = Digital Versatile Disc,
  tag = abbrev
}
\DeclareAcronym{avc}{
  short = AVC ,
  long  = Advanced Video Coding,
  tag = abbrev
}
\DeclareAcronym{aac}{
  short = AAC ,
  long  = Advanced Audio Coding,
  tag = abbrev
}
\DeclareAcronym{mpeg}{
  short = MPEG ,
  long  = Moving Picture Experts Group,
  tag = abbrev
}
\DeclareAcronym{jctvc}{
  short = JCTVC ,
  long  = Joint Collaborative Team on Video Coding,
  tag = abbrev
}

\DeclareAcronym{itut}{
  short = ITU-T ,
  long  = International Telecommunication Union- Telecommunications standardization,
  tag = abbrev
}
\DeclareAcronym{vceg}{
  short = VCEG ,
  long  = Video Coding Experts Group,
  tag = abbrev
}
\DeclareAcronym{iso}{
  short = ISO ,
  long  = International Organization for Standardization,
  tag = abbrev
}
\DeclareAcronym{iec}{
  short = IEC ,
  long  = International Electrotechnical Commission,
  tag = abbrev
}

\DeclareAcronym{cu}{
  short = CU ,
  long  = Coding Unit,
  tag = abbrev
}
\DeclareAcronym{pu}{
  short = PU ,
  long  = Prediction Unit,
  tag = abbrev
}
\DeclareAcronym{cb}{
  short = CB ,
  long  = Coding Block,
  tag = abbrev
}
\DeclareAcronym{tb}{
  short = TB ,
  long  = Transform Block,
  tag = abbrev
}

\DeclareAcronym{ctb}{
  short = CTB ,
  long  = Coding Tree Block,
  tag = abbrev
}
\DeclareAcronym{ctu}{
  short = CTU ,
  long  = Coding Tree Unit,
  tag = abbrev
}

\DeclareAcronym{amp}{
  short = AMP ,
  long  = Advanced Motion Prediction,
  tag = abbrev
}
\DeclareAcronym{rdo}{
  short = RDO ,
  long  = Rate-Distortion Optimization,
  tag = abbrev
}
\DeclareAcronym{rqt}{
  short = RQT ,
  long  = Residual QuadTree,
  tag = abbrev
}

\DeclareAcronym{mv}{
  short = MV ,
  long  = Motion Vector,
  tag = abbrev
}
\DeclareAcronym{cabac}{
  short = CABAC ,
  long  = Context-Adaptive Binary Arithmetic Coding,
  tag = abbrev
}
\DeclareAcronym{rd}{
  short = R-D ,
  long  = Rate-Distortion,
  tag = abbrev
}

\DeclareAcronym{qp}{
  short = QP ,
  long  = Quantization Parameter,
  tag = abbrev
}
\DeclareAcronym{bdrate}{
  short = BD-Rate ,
  long  = Bjøntegaard Delta Rate,
  tag = abbrev
}
\DeclareAcronym{bdpsnr}{
  short = BD-PSNR ,
  long  = Bjøntegaard Delta Peak Signal to Noise Ratio,
  tag = abbrev
}

\DeclareAcronym{simd}{
  short = SIMD ,
  long  = Single Instruction Multiple Data,
  tag = abbrev
}
\DeclareAcronym{isa}{
  short = ISA ,
  long  = Instruction Set Architecture,
  tag = abbrev
}
\DeclareAcronym{jvet}{
  short = JVET ,
  long  = Joint Video Exploratory Team,
  tag = abbrev
}

\DeclareAcronym{sse}{
  short = SSE ,
  long  = Streaming SIMD Extensions,
  tag = abbrev
}
\DeclareAcronym{sse2}{
  short = SSE2 ,
  long  = Streaming SIMD Extensions 2,
  tag = abbrev
}
\DeclareAcronym{ssse3}{
  short = SSSE3 ,
  long  = Supplemental Streaming SIMD Extensions 3,
  tag = abbrev
}
\DeclareAcronym{sse4}{
  short = SSE4 ,
  long  = Streaming SIMD Extensions 4,
  tag = abbrev
}

\DeclareAcronym{avx}{
  short = AVX ,
  long  = Advanced Vector Extensions,
  tag = abbrev
}
\DeclareAcronym{avx2}{
  short = AVX2 ,
  long  = Advanced Vector Extensions 2,
  tag = abbrev
}
\DeclareAcronym{avx512}{
  short = AVX512 ,
  long  = Advanced Vector Extensions 512,
  tag = abbrev
}
\DeclareAcronym{fma}{
  short = FMA ,
  long  = Fused Multiply-Accumulate,
  tag = abbrev
}

\begin{document}

\frontmatter

\titlepage

\begin{abstract}
Multi-encoding implies encoding the same content in multiple spatial resolutions and multiple bitrates. This work evaluates the encoder analysis correlations across 2160p, 1080p, and 540p encodings of the same video for conventional \ac{abr} bitrates. A multi-resolution tier multi-\ac{abr} encoding scheme is modeled and evaluated, which significantly improves the computational efficiency of conventional \ac{abr} encoding. Video content is first encoded at the lower resolution with the associated median bitrate, and encoder analysis decisions, such as motion vectors and \ac{cu} block structure, are then used in the other encodes in the same resolution tier. The analysis is then extrapolated and refined to be used in higher-resolution encodes. The scheme is validated using x265 \ac{hevc} video encoder. The proposed multi-resolution tier multi-bitrate encoding scheme achieves overall speed-ups of up to 2.5x, compared to the conventional single-instance encoding approach. Furthermore, this speed-up is achieved without substantial losses in coding efficiency.\\

\noindent \ac{simd} Vector units in \ac{cpu}s have become the de-facto standard for accelerating media and other kernels that exhibit parallelism. This work also demonstrates the impact of hardware-aware optimizations on the encoding speeds of the next-generation video codecs. The work is evaluated using the Arowana XVC encoder.
\end{abstract}

\begin{otherlanguage}{swedish}
  \begin{abstract}
Att komprimera samma videosekvens i olika upplösningar och vid olika bithastighet kan kallas "multi-encoding". Denna uppsats utvärderar komprimeringsanalys vid upplösning 2160p, 1080p och 540p av samma videosekvens med bithastighet anpassad för adaptiv videoströmming. En komprimeringsmetod med tre nivåer som ger en signifikant förbättring i beräkningsprestanda har modellerats och utvärderts. Videosekvenserna komprimeras först vid lägsta upplösningen och kodningsval såsom rörelsevektorer och blockstuktur används sedan vid komprimering i samma upplösining och även i högre upplösning genom extrapolation. Modellen har validerats genom att använda x265 \ac{hevc}-kodaren. Den föreslagna modellen med tre nivåer ger en sammanlagd uppsnabbning om 2.5x i genomsnitt, jämfört med konventionell komprimering. Uppsnabbningen åstadkoms utan substantiell förlust i kodningseffektivitet.\\

\noindent Moderna processorers \ac{simd} instruktioner har blivit de-facto standard för acceleration av mediaprocessesning och liknande beräkningar som kan utnyttja parallelism. Denna uppsats påvisar också effekten av hådvaruoptimering för uppsnabbning av nästa generations video codecs. Arbetet är utfört i videokodaren Arowana XVC.
  \end{abstract}
\end{otherlanguage}

\tableofcontents
\listoffigures

\listoftables
\newpage
\printacronyms[include=abbrev, name=List of abbreviations]

\mainmatter

\include{kth_acknowledgement}
\include{kth_chapter1_introduction}

\include{kth_chapter2_hevc_abr_intro}
\include{kth_chapter3_xvc_simd_intro}
\include{kth_chapter4_analysis_methods}
\include{kth_chapter5_simd_methods}
\include{kth_chapter6_results}
\include{kth_chapter7_conclusion}

\printbibliography[heading=bibintoc]

\appendix
\include{kth_appendix1_x265_options}
\include{kth_appendix2_sse4}
\include{kth_appendix3_avx2}

\tailmatter

\end{document}

%% file: kth_acknowledgement.tex
\chapter*{Acknowledgements}

I am deeply indebted to my supervisor \textbf{Mr. Jonatan Samuelsson}, CEO, Divideon, and my examiner \textbf{Dr. Markus Flierl}, Electrical Engineering and Computer Science, KTH Royal Institute of Technology, Sweden, for their continual support, ardent motivation, and enduring guidance throughout my project tenure.\\

\noindent I further extend my gratitude to Dr. Sundararaman Gopalan, Mrs. Gayathri N, Department of Electronics and Communication, Amrita School of Engineering Amritapuri Campus, India, my close friends who had worked with me closely in MulticoreWare Inc namely Kalyan Goswami, Praveen Tiwari, Santhoshini Sekar, Bhavna Hariharan and Kavitha Sampath for their belief in me and helping me in growing my learning curve.\\

\noindent Finally, I express my gratefulness to my parents, my uncle Dr. Venugopalan P, my aunt Mrs. Jyothi PM, and my sister Aishwarya for their love, encouragement, and support. I also acknowledge the support and motivation of my three dear friends in Stockholm- Anubhab Ghosh, Prajit T Rajendran and Shreya K Chari.

%% file: kth_chapter1_introduction.tex
\chapter{Introduction}
\begin{minipage}{0.65\textwidth}
There is a tremendous increase in video content on streaming services over the last few years. The figure on the right shows the application-wise global internet traffic, where 62 \% comprises video streaming. According to The Global Internet Phenomena Report from Sandwine, video plays in March 2020 have doubled compared to the end of February 2020 during the COVID-19 pandemic \cite{intro_ref}. Hundreds of millions of subscribers and transactional \ac{vod} users are part of the internet traffic. The amount of available live linear and on-demand assets in \ac{uhd} \cite{uhd_ref} is significantly growing. E.g., more than 250 cinematic titles are now available from major studios in \ac{uhd} \ac{hdr}, and it is noteworthy that FIFA World Cup 2018 was streamed live in \ac{uhd} \ac{hdr} \cite{hevc_ref1}. The situation calls for effective video compression schemes to reduce the global network load. 
\end{minipage}
\hfill
\begin{minipage}{0.28\textwidth}
\includegraphics[width=\textwidth]{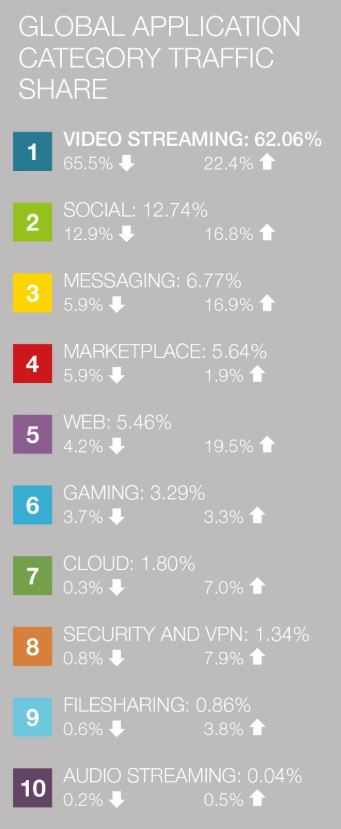}
Global internet traffic. Source: Sandvine \cite{intro_ref}
\end{minipage}

\section{Introduction to video compression}
Video encoding refers to the process of converting raw video into a digital format that’s compatible with many devices. This is for compatibility and efficiency with a desired set of applications and hardware such as for \ac{dvd}/Blu-ray, mobile, video streaming, or general video editing. The encoding process transforms the video and audio data into a file and compresses according to the chosen encoding standard's specifications.\\

\noindent To shrink a video into a more manageable size, content distributors use a video compression technology called a \textit{codec}. Codecs allow us to compress a bulky video for delivery and storage tightly. Codecs apply algorithms to the video and create a facsimile of it. The video is shrunk down for storage and transmission and later decompressed for viewing. Streaming employs both audio and video codecs. H.264, also known as \ac{avc}, is the most common video codec. \ac{aac} is the most common audio codec.\\

\noindent Once compressed, the components of a stream are packaged into a wrapper or file format. These files contain the audio codec, video codec, closed captioning, and associated metadata. Common containers include .mp4, .mov, .ts, and .wmv. Containers can often input multiple types of codecs. That said, not all playback platforms accept all containers and codecs. That’s why multi-format encoding is crucial when streaming to many devices.\\

\noindent In current world of video compression, \ac{uhd} and \ac{hdr} content calls for using the \ac{hevc} video compression standard \cite{hevc_main_ref}. Other codecs are being developed to mitigate the royalty issues associated with \ac{hevc}- XVC codec is one of them. As the codecs developed for \ac{uhd} video content comprises a significant amount of computations to achieve the desired compression efficiency, there is always scope for optimization of the encoding pipeline to achieve the best trade-off between computational speed, compression efficiency and visual quality.\\

\noindent On the other hand, the adaptive streaming approach \cite{abr_ref1} is nearly universally used over the last several years to reach heterogeneous devices on non-provisioned networks (such as the Internet). The key property of adaptive streaming from the standpoint of this project is that the same video content is made available to the player in different bitrates and resolutions (“bitrate ladder”). Encoding all steps of a bitrate ladder for 10-bit \ac{hevc} content requires a significant amount of computational resources – e.g. sub-1 frames per second (fps) encoding speeds are very common for high-quality offline encoding. Section \ref{sec:hevc_background} and Section \ref{sec:ABR_background} provide a more detailed introduction and background study to \ac{hevc} as well as to \ac{abr} streaming, respectively.

\section{Introduction to Single Instruction Multiple Data}
Over the years, tremendous advancements have been seen in computer architecture. This field also has a huge influence in the areas of video coding, especially in \ac{simd} instructions deployment. The category of a parallel computer in Flynn’s taxonomy constitutes \ac{simd}. Computers that perform the same operation simultaneously on multiple data points are described here. They also have multiple processing elements. The development of software only \ac{mpeg}-1 with general-purpose processors that can do real-time decoding is the reason for the introduction of \ac{simd} extensions \cite{simd_ref_z}\cite{simd_ref_y}. Adjustment of the volume of digital audio and contrast in a digital image are the common applications of \ac{simd} extensions. \ac{simd} instructions are also utilized in modern \ac{cpu} designs for enhancing multimedia use performance. Increasing the efficiency of video coding and standardization are other use cases of \ac{simd} capabilities. This minimizes the effort in the elimination of sample dependencies and definition of computation precision.\\

\noindent \ac{isa}s, along with their micro-architectures, are getting more diversified in recent years. There is also an increase in complexity, like the inclusion of multiple instruction sets for \ac{simd} \ac{isa}s. In addition, instructions have various performance features  even after using the same \ac{isa}. Video codecs these days have the significant capability for the acceleration of \ac{simd}, which intensifies the need for more investigation on the effect of these accelerations on video encoders. \ac{avx512}, which is the new \ac{simd} extension released by Intel, are 512-bit extensions to the previous 256-bit Vector Extensions \ac{simd} instructions for x86 \ac{isa}. \ac{avx512} can accumulate eight 64-bit integers, eight double-precision, sixteen 32-bit integers, or sixteen single-precision floating point numbers within 512-bit vectors. This is around four times of \ac{sse} and twice the amount of data points \ac{avx}/ \ac{avx2} can compute with one instruction.\\

\noindent This project aims to focus on the \ac{simd} impact on video codec- XVC. The research tries to estimate the dependency of the complexity of the codec on the performance benefits achieved from \ac{simd} extensions. The scope of research and analysis in this project is limited to Intel \ac{avx2} \ac{simd} and Intel \ac{sse4} instructions, along with their potential benefits in the encoder speeds.

\section{Report Structure}
Chapter \ref{sec:hevc_abr_background} gives brief insights on the background of \ac{hevc} \ac{abr} streaming. Section \ref{sec:hevc_background} gives details on the background of \ac{hevc}- its establishment as a leading video codec and some key technical details. Section \ref{sec:x265_background} gives a brief background of x265, an open source \ac{hevc} encoder used in this work to evaluate the encoding approaches. Section \ref{sec:ABR_background} describes the \ac{abr} streaming terminology- its development and some key concepts. Section \ref{sec:abr_related_work} mentions some of the previous works in the domain of this work, which are used in the literature study of this work.\\

\noindent Chapter \ref{sec:xvc_simd_background} gives brief insights on the XVC video codec and Intel \ac{simd} optimization. Section \ref{sec:xvc} discusses the background of XVC video codec, while Section \ref{sec:simd_sec} explains the background of \ac{simd} extensions. Section \ref{sec:sse4} and Section \ref{sec:avx2} introduce Intel \ac{sse4} and \ac{avx2} \ac{simd} extensions. Section \ref{sec:lit_survey_simd} presents some of the literature referred to for this project.\\

\noindent Chapter \ref{sec:analysis_share_x265_method} gives brief insights on the methods used in this project for multi-resolution and multi-rate analysis sharing in x265. Section \ref{sec:testing_method} shows how the various methods considered in this project are tested. Section \ref{sec:intra_res_share_x265} explains the methods used in intra-resolution analysis sharing where Section \ref{sec:intra_res_stateofart}  details the state of the art method while Section \ref{sec:intra_analysis_median} describes the proposed method. Section \ref{sec:inter_res_share_x265} shows the details of the method used in Inter resolution analysis sharing.\\

\noindent Chapter \ref{sec:xvc_simd_method} presents the work carried out to understand the impact of \ac{simd} parallelism in media processing. Section \ref{sec:simd_impl_methods} shows various methods used for \ac{simd} implementation and motivates the choice of the hand-written assembly method. Section \ref{sec:nasm} explains about NASM assembler used for compiling hand-written assembly code in this project. Section \ref{sec:acc_xvc_simd} shows the \ac{cpu} cycle count improvements for kernels accelerated with Intel \ac{sse4} and \ac{avx2} \ac{simd}. Section \ref{sec:sel_kernel} mentions how the functions to be optimized are chosen, Section \ref{sec:cost_func} and Section \ref{sec:simd_cost_func} describing the implementation of encoder cost functions and the resulting \ac{cpu} cycle count improvements. Section \ref{sec:cycl_improve_method} explains the method used to evaluate the \ac{cpu} cycle count improvements for each optimized function.\\

\noindent Chapter \ref{sec:Results} evaluates the methods described in Section \ref{sec:analysis_share_x265_method} in the real time streaming scenarios. Section \ref{sec:intra_abr} shows the state-of-the-art intra-resolution analysis sharing method in \ac{cvbr} mode and discusses the associated results. Section \ref{sec:novel_intra_abr} discusses the proposed intra-resolution analysis sharing method in \ac{cvbr} mode and its improvement over the state-of-the-art intra-resolution analysis sharing method. Section \ref{sec:inter_abr} explains the inter-resolution analysis sharing method in \ac{cvbr}, the need for analysis refinement, and the associated results. Section \ref{sec:stateofart_multi_res_rate} and Section \ref{sec:proposed_multi_res_rate} discuss state-of-the-art and the proposed multi-encoder structure with the scheme for encoder analysis sharing and the associated results, respectively. Section \ref{sec:xvc_simd_res} discusses the improvements of the Arowana XVC encoding speeds utilizing the \ac{simd} extensions.\\

\noindent Chapter \ref{sec:conc_future} concludes the project report and summarises the key findings during the project. Section \ref{sec:conclusion} discusses the conclusions and Section \ref{sec:future} discusses the future directions of this work which haven't been explored as part of this project respectively.

%% file: kth_chapter2_hevc_abr_intro.tex
\chapter{HEVC Adaptive Bitrate Streaming}
\label{sec:hevc_abr_background}
This chapter gives brief insights into the background of \ac{hevc} \ac{abr} streaming. Section \ref{sec:hevc_background} gives details on the background of \ac{hevc}- its establishment as a leading video codec and some key technical details. Section \ref{sec:x265_background} gives a brief background of x265, an open source \ac{hevc} encoder used in this work to evaluate the encoding approaches. Section \ref{sec:ABR_background} describes the \ac{abr} streaming terminology- its development and some key concepts. Section \ref{sec:abr_related_work} mentions some of the previous works in the domain of this work, which are used in the literature study of this work.

\section{\ac{hevc} Background}
\label{sec:hevc_background}
\ac{hevc} was finally endorsed as an encoding standard as the successor of the hugely popular \ac{avc} standard by the \ac{jctvc} \cite{hevc_main_ref}. Its inception was carried out over 2.5 years, and the first version was finalized in 2013 \cite{xvc_5}. The \ac{jctvc} was created originally by a collaboration of  \ac{itut} \ac{vceg} and \ac{iso} \ac{iec} \ac{mpeg} working on the \ac{hevc} standard. The final \ac{hevc} specification was approved as part of the Recommendation H.265 by \ac{itut} and as \ac{mpeg}.H Part 2 by \ac{iso} \ac{iec}. The most recent version of \ac{hevc} in use is its 7th version \cite{hevc_5}. \ac{hevc} development emphasized performing efficient encoding of high-resolution videos, which leads to a significant coding gain while encoding \ac{uhd} (4K or 2160p resolution) videos \cite{uhd_ref}. This is accomplished in the \ac{hevc} standard by incorporating tools for increasing coding efficiency.\\

\noindent The basic processing units in \ac{hevc} are created by partitioning each video frame into square-shaped \ac{ctb}s. \ac{ctb}s come in varying sizes such as 16x16, 32x32 or 64x64. A more complex structure called \ac{ctu} is created by combining the \ac{ctb} with associated syntax elements: one luma and two corresponding chroma. By increasing the size of the \ac{ctu} we can obtain a better coding efficiency at high resolutions at the cost of an increase in computational complexity. \ac{hevc} uses a hierarchical quad-tree partitioning structure which splits the \ac{ctu} into one more of coding units or \ac{cu}s of varying sizes between 8x8 and 64x64. In addition, it is also possible to subdivide the \ac{cu}s into smaller blocks along coding tree boundaries for intra-picture (spatial) and inter-picture (temporal motion compensated) predictions. If the candidate block is present in the same frame, it is termed intra-predicted, whereas if it is from a different frame, it is said to be inter-predicted. Intra-predicted blocks can be represented as a combination of a prediction block and a mode denoting the angle of prediction. The modes used in intra-prediction are DC, planar and angular modes. They represent various angles from the predicted block. On the other hand, inter-predicted blocks can be represented as a combination of the reference block (block used for prediction) and the \ac{mv}, which represents the delta difference between the reference block and the current block. The block with a net zero \ac{mv} uses the merge mode, and the others use the \ac{amp} mode. In the special case where there is no residual, ie. the predicted block is identical to the source, we use the skip mode- a special case of the merge mode. To provide the prediction data, a minimum of one \ac{pu} is defined for each \ac{cu}. The prediction mode selected indicates whether the \ac{cu} (comprising of the \ac{cb}, single luma, and two chroma) is coded in the intra-picture prediction mode or the inter-picture prediction mode. \ac{tb}s can be formed by partitioning a \ac{cb} in size varying from 4x4 to 32x32 to perform transform coding of the prediction residuals. Each \ac{ctb} thus acts as a root node of the coding tree, and the coding block acts as the leaf of the coding tree. The coding block is a root of the transform or \ac{rqt}.\\

\begin{figure}[!ht]
    \centering
    \includegraphics[width=0.90\textwidth]{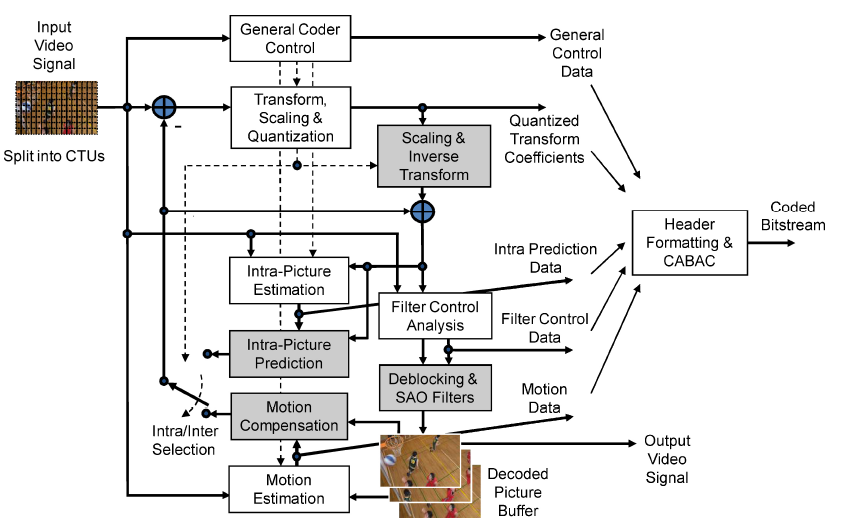}
    \caption{HEVC Encoder block diagram \cite{hevc_pic_ref}}
    \label{fig:hevc_enc_diagram}
\end{figure}

\noindent \ac{rdo} \cite{rdo_ref} is used in intra or inter-picture prediction, including the quad-tree and RQT partitioning. \ac{rdo} aims to ensure that distortion is minimized at the pre-set target bit rate or that the bit rate is minimized at a target quality level for distortion pre-set. In a high-resolution video, efficient encoding would require many block-splitting decisions and, thereby, a large number of \ac{rdo} calculations. This translates to a higher drain on computational resources. Reducing the overall computational complexity of encoding is thus essential in providing a high quality \ac{uhd} visual experience to a large audience economically.\\ 

\noindent After the finalization of the \ac{hevc} standard in 2013, various open-source implementations for video encoding into the H.265/\ac{mpeg}-\ac{hevc} compression format have come to the fore- x265 is one of them. The background on x265 is explained in Section \ref{sec:x265_background}.

\section{x265 Background}
\label{sec:x265_background}
The x265 is an open-source \ac{hevc} encoder that compresses raw video in compliance with the \ac{hevc} standard \cite{hevc_main_ref}. This encoder is integrated into several open-source frameworks, including VLC, HandBrake \cite{hb_ref}, and FFMpeg \cite{ff_ref} and is the de-facto open-source video encoder for \ac{hevc}. The x265 encoder has assembly optimizations for several platforms, including Intel architecture, ARM, and PowerPC. The x265 encoder employs techniques for inter-frame and intra-frame parallelism to deal with the increased complexity of \ac{hevc} encoding. For inter-frame parallelism, x265 encodes multiple frames in parallel by using system-level software threads. For intra-frame parallelism, x265 relies on the Wave-front Parallel Processing (WPP) tool exposed by the \ac{hevc} standard. This feature enables encoding rows of \ac{ctu}s of a given frame in parallel while ensuring that the blocks required for intra-prediction from the previous row are completed before the given block starts to encode; as per the standard, this translates to ensuring that the next \ac{ctu} on the previous row completes before starting the encode of a \ac{ctu} on the current row. The combination of these features gives a tremendous boost in speed with no loss in efficiency compared to the publicly available reference encoder, HM.

\section{\ac{abr} Streaming Background}
\label{sec:ABR_background}

\ac{abr} is a technology that allows streaming files efficiently over \ac{http} networks. The user's video player is offered multiple files on the same content in different sizes, and the client selects the most appropriate file for playback on the device. The design focuses on improving streaming by delivering the right content in all circumstances based on the specific device and network condition, which decreases the need for re-buffering. \ac{abr} streaming makes it possible for video players to select the best available video segment or chunk to play, depending on the available bitrate and device capability.\\

\noindent The \ac{abr} Ladder is an array of segments of differing quality and resolution, which are available in the streaming server. If the bitrate increases, the video player logic chooses a file of larger size with better quality, analogous to climbing up the ladder. If the bitrate decreases, the video player switches back to a lower-quality file, analogous to climbing down the ladder. \ac{abr} streaming improves previous streaming models by adjusting the stream to choose the most suitable bitrate and changing the transport conditions based on the current situation. This makes it extremely useful for streaming on mobile networks. \ac{abr} streaming uses dynamic tracking of parameters such as \ac{cpu},  memory capacity, and network conditions and subsequently delivers video quality to match.\\ 

\noindent \ac{abr} is supported by most modern video players. The video player logic on the user's device can choose among all the segments available in the video's manifest file, which are adjusted to increase the bitrates. The video player selects the best match based on the available bandwidth on the user's device at a particular moment. The player initially requests the lowest bit rate segments that are on offer. If the video player determines that the download speed is higher than the bitrate capability of the current segment, it will send a request for the next higher bitrate segment. This process continues until the current bitrate segment and the available bandwidth is matched appropriately. The video player will continue requesting segments at this selected bandwidth as long as the bandwidth remains at a similar level. If this process works smoothly, the user would have a smooth viewing experience in different network conditions. \\

\begin{table}[!htbp]
\centering
\begin{tabular}{|p{2.5cm} | p{2.0cm} | p{1.5cm}| p{1.5cm} |p{1.8cm} | p{1.8cm} |}
\hline
Network Type & Dimensions & Frame rate & Total Bitrate & Audio bitrate & Key Frame\\
\hline
\rowcolor{gray!10} Cell &	480x320 &	N/A &	64 & 64  & N/A \\
\rowcolor{blue!5} Cell &	416x234 &	10-12 &	264 & 64 & 30-64 \\
\rowcolor{gray!10} Cell &	480x270 &	10-12 &	464 & 64 & 90  \\
\rowcolor{blue!5} WiFi &	640x360 &	29.97 &	664 & 64 & 90 \\
\rowcolor{gray!10} WiFi &	640x360 &	29.97 &	1264 & 64 & 90  \\
\rowcolor{blue!5} WiFi &	960x540 &	29.97 &	1864 & 64 & 90 \\
\rowcolor{gray!10} WiFi &	960x540 &	29.97 &	2564 & 64 & 90  \\
\rowcolor{blue!5} WiFi &	1280x720 &	29.97 &	4564 & 64 & 90 \\
\rowcolor{gray!10} WiFi &	1280x720 &	29.97 &	6564 & 64 & 90  \\
\rowcolor{blue!5} WiFi &	1920x1080 &	29.97 &	8564 & 64 & 90 \\
\hline
\end{tabular}
\caption{Fixed Bitrate Encoding ladder proposed by Apple in Apple Tech Note TN2224 \cite{abr_ref1}}
\label{tab:abr_ref_tab1}
\end{table}

\noindent Apple proposed a Fixed Bitrate Encoding ladder in 2010, shown in Table \ref{tab:abr_ref_tab1}. Here, a video file would be encoded into variants of ten distinct qualities, ranging from an audio-only stream at 64 kbps to a 1080p audio/video stream at 8564 kbps. The Apple dimension numbers reported for cell phones are for portrait review, whereas those reported for Wi-Fi are for landscape view. \\

\noindent Netflix refined the technique of \ac{abr} Ladder analysis with Per-Title Encoding in 2015. In this particular approach by Netflix, each video is encoded in multiple resolutions. The data rates identify the appropriate rungs that provide the best quality at each data rate, respectively, as shown in Table \ref{tab:abr_ref_tab2}. This ensured that a content provider could customize two parameters- the number of rungs in a ladder, and the related resolution based on the requirements for each specific video, as shown below. This strategy provided potential savings in encoding and storage. 

\begin{table}[!htbp]
\centering
\begin{tabular}{|p{2.4cm} | p{4.7cm} | p{4.5cm}|}
\hline
Resolution & Fixed Bitrate ladder (kbps) & Per-title Ladder (kbps) \\
\hline
\rowcolor{gray!10} 320x240 &	235 &	150 \\
\rowcolor{blue!5} 384x288 &	375 &	200 \\
\rowcolor{gray!10} 512x384 & 560 &	290  \\
\rowcolor{blue!5} 512x384 &	750 &	NA \\
\rowcolor{gray!10} 640x480 & 1050 &	NA  \\
\rowcolor{blue!5} 720x480 &	1750 &	440 \\
\rowcolor{gray!10} 720x480 & NA  &	590  \\
\rowcolor{blue!5} 1280x720 & 2350 &	830 \\
\rowcolor{gray!10} 1280x720 & 3000 &	1150  \\
\rowcolor{blue!5} 1920x1080 & 4300 &	1470 \\
\rowcolor{gray!10} 1920x1080 & 5800 &	2150 \\
\rowcolor{blue!5} 1920x1080 & NA &	3840 \\
\hline
\end{tabular}
\caption{Per-Title Encoding proposed by Netflix \cite{abr_ref1}}
\label{tab:abr_ref_tab2}
\end{table}

\begin{figure}[!ht]
    \centering
    \includegraphics[width=0.55\textwidth]{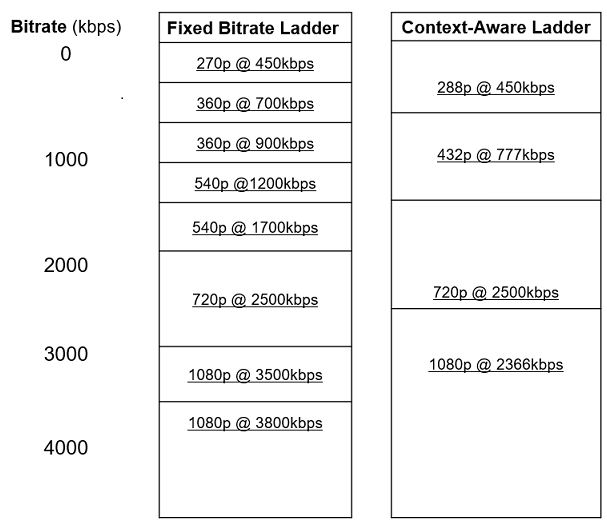}
    \caption{Context-Aware Encoding \cite{abr_ref1}}
    \label{fig:abr_ref_pic3}
\end{figure}

\noindent Context-Aware Encoding (CAE) started gaining traction around 2018. It expanded encoding considerations also to include devices. Using Context-Aware Encoding, a content provider could offer separate encoding parameters for each device type and deliver content to various device types, from smartphones to home theatres. The aim is to decrease the playback bandwidth while maintaining the same experience. This reduction in playback bandwidth can be quite substantial. The Context-Aware Ladder shown in Figure \ref{fig:abr_ref_pic3} can offer the same quality as the Fixed Bitrate Ladder by using lower bitrates or higher resolutions for each variant, thereby improving playback performance and cost efficiency, and with half the number of variants. If encoding is aimed at saving the overall bandwidth, one important factor would be the location where the content will be consumed. The traditional encoding approach involves creating typical settings for each video in a library. For situations wherein the video will be streamed to multiple devices, such as mobile devices on cellular data networks and wired set-top boxes, it is advisable to have multiple encoding schemes, one for each type of device expected. Context-Aware Encoding ensures that content providers can fine-tune parameters on a per-show or a per-scene encoding basis.\\

\noindent The critical aspect of adaptive streaming relevant to this work is that multiple representations are needed for a broader reach. This work proposes a method to decrease the computational redundancy in encoding the same video content at multiple resolutions and bitrates for adaptive streaming distribution.   

\section{Related Work}
\label{sec:abr_related_work}

The work \cite{reuse_pap_4} proposes an approach wherein the highest bitrate/ quality representation is encoded with unmodified \ac{hevc}. The dependent representations are then encoded based on the \ac{cu} depth information. Since \ac{cu}s typically tend to have higher depth values (smaller \ac{cu} size) in higher quality by observation, we only search \ac{cu}s for which their depth levels are smaller or equal to that of the co-located \ac{cu} in the reference representation. Using this approach, larger depths are skipped in the search, and we can save a considerable amount of time.\\

\noindent Following this idea, \cite{reuse_pap_5} discusses a similar approach using representations with different resolutions. Since the \ac{ctu}s in different resolutions encompass different corresponding areas, they do not match with each other. Thus, a novel matching algorithm is proposed here with the objective to obtain block structure information from high-resolution representation with an arbitrary down-sampling ratio. This information is subsequently used to skip unnecessary depth searches at the lower resolutions. The two methods mentioned above are combined to introduce an efficient multi-rate encoding method in the work discussed in \cite{reuse_pap_3}. Here, in addition to the \ac{ctu} depth values, additional information from the reference representation- namely prediction mode, intra mode, and motion vectors are used to accelerate encoding of lower quality and lower resolution representations.\\

\noindent The approaches discussed above, make use of re-utilization of high-resolution results at the lower “dependent” or derived resolutions. This re-utilization results in a lesser usage requirement of computational resources as encoding low-resolution content is inherently faster. The work \cite{reuse_pap_2} proposed an approach with sharing analysis information across representations in x265 - but here, all representations are restricted to the same resolution.\\

\begin{figure}[!ht]
    \centering
    \includegraphics[width=0.60\textwidth]{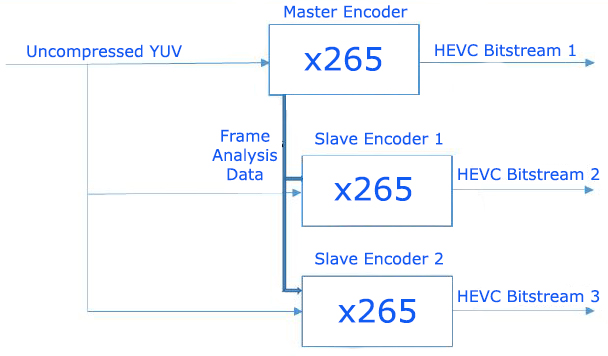}
    \caption{Multi-rate implementation proposed in \cite{reuse_pap_3}, \cite{reuse_pap_2} }
    \label{fig:multirate_imp_x265_diag}
\end{figure}

\noindent \cite{reuse_pap_6} introduces a novel implementation of the x265-based intelligent framework for multi-resolution encoding. Using such a framework, the relevant information related to the encoding decisions, such as a quad-tree structure, prediction modes, etc., which are made during the encode, can be shared from the lower resolution encode to the higher resolution encode. This makes the higher-resolution encoding process by up to two times faster.

\begin{figure}[!ht]
    \centering
    \includegraphics[width=0.70\textwidth]{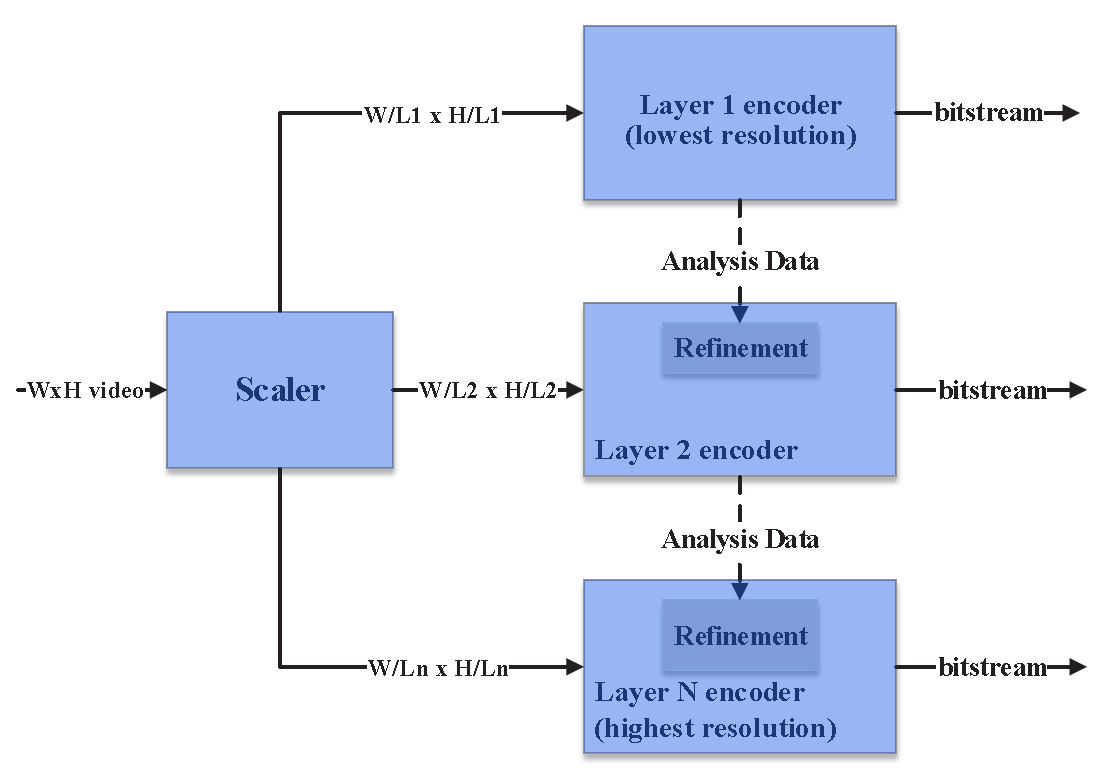}
    \caption{Multi-rate implementation proposed in \cite{reuse_pap_6}}
    \label{fig:multires_imp_x265_diag}
\end{figure}

%% file: kth_chapter3_xvc_simd_intro.tex
\chapter{XVC and Intel SIMD Background}
\label{sec:xvc_simd_background}
This chapter gives brief insights on the XVC video codec and Intel \ac{simd} optimization. Section \ref{sec:xvc} discusses the background of XVC video codec, while Section \ref{sec:simd_sec} explains the background of \ac{simd} extensions. Section \ref{sec:sse4} and Section \ref{sec:avx2} introduce Intel \ac{sse4} and \ac{avx2} \ac{simd} extensions. Section \ref{sec:lit_survey_simd} presents some of the literature referred to for this project. 

\section{XVC video codec}
\label{sec:xvc}
The video codec named XVC is developed by the software video compression company named Divideon AB and the first version of it was released in September 2017 \cite{xvc_intro_ref}. It has been developed mainly based on known technology that has been included in AVC \cite{xvc_4} or \ac{hevc} \cite{hevc_main_ref}, or that has been evaluated in the context of the \ac{jvet} \cite{xvc_6}. The XVC video codec almost offers a middle-way between the existing options represented by \ac{hevc} and AV1 \cite{av1_ref}. It is a block-based codec using intra and inter-predictions that operates on raw pictures of YUV pixel samples into a pre-defined standard bit-stream. Each picture in a video sequence is divided into rectangular blocks of samples of sizes up to 64x64 samples, which are predicted from samples in the same picture (intra-prediction) or samples in previously coded pictures (inter-prediction). Residuals are transformed using non-square transforms, and the coded symbols are compressed using a \ac{cabac} coder. Block boundaries are filtered using a deblocking filter.\\

\noindent In general, XVC is more similar to \ac{hevc}, but there are several clear differences. One of the most significant differences is that XVC uses non-square coding units for which both prediction and transform are applied. Thus, XVC does not contain separate trees of prediction units and transform units, as is the case with \ac{hevc}. It includes extensions to technologies in \ac{hevc} in several areas, for example, 67 intra-prediction modes instead of 35. Still, there is also a significant number of new coding tools for which there is no corresponding technology in \ac{hevc}, such as:
\begin{itemize}
    \item Adaptive motion vector precision – where the precision of the \ac{mv}s are signaled to allow for more efficient signaling of long and integer \ac{mv}s.
\item Affine motion prediction – where individual \ac{mv}s are calculated and applied for each 4x4 sub-block of a \ac{cu} which makes it possible better to represent non-translational motions such as rotation and zoom.
\item  Cross component prediction – where chroma samples are predicted from luma samples using a linear model.
\item Transform selection – where different transforms with different characteristics are evaluated to determine which transform most efficiently represents the residual of a specific \ac{cu}.
\item Local illumination compensation – where a linear model accounts for local offsets of sample values when predicting from reference pictures, particularly useful for representing changes in lighting conditions of an object or a scene.
\end{itemize}

\noindent The XVC codec has been tested with several open-source applications, including ExoPlayer, FFmpeg, and VLC. These integrations have made it possible to use XVC on various devices and platforms, including Android, iOS, Windows, and Linux.

\section{Single Instruction Multiple Data extension}
\label{sec:simd_sec}
\ac{simd} is a class of parallel computers in Flynn’s taxonomy \cite{Flynn2011}. Program with \ac{simd} can simultaneously perform the same operation for multiple data elements. This is data-level parallelism, not concurrency, which means multiple data elements are being processed simultaneously, but only one operation is performed. The program, which can potentially run in data-level parallelism, can be significantly accelerated by \ac{simd}.\\

\noindent The concept of \ac{simd} was introduced in the early 1970s, known as a "vector" of data with single instructions, but it is separated from \ac{simd} now. Afterward, the \ac{simd} was introduced in modern \ac{simd} machines which has many limited-functionality processors. Then the domain of \ac{simd} \ac{cpu} changed from desktop-computer to supercomputer. Since the desktop computer started to become performance efficient to support real-time gaming and audio/ video processing during the 1990s, the \ac{simd} extensions went back to the desktop \ac{cpu}s again.\\

\begin{table}[!htbp]
\centering
\begin{tabular}{ |p{3.8cm}| p{2.8cm}| p{1.8cm}| p{2.4cm}| p{1.8cm}|}
\hline
\ac{simd} \ac{isa} & Base \ac{isa} & Vendor & Year & \ac{simd} Registers\\
\hline
\rowcolor{gray!10} MMX & x86 & Intel &  1996 & 8x64b \\
\rowcolor{blue!5} 3DNow! & x86 & AMD & 1998 & 8x64b \\
\rowcolor{gray!10} SSE & x86/x86-64 & Intel & 1999 & 8/16x28b \\
\rowcolor{blue!5} SSE2 & x86/x86-64 & Intel & 2000 & 8/16x28b \\
\rowcolor{gray!10} SSE3 & x86/x86-64 & Intel & 2004 & 8/16x28b \\
\rowcolor{blue!5} SSSE3 & x86/x86-64 & Intel & 2006 & 8/16x128b \\
\rowcolor{gray!10} SSE4 & x86/x86-64 & Intel & 2007 & 8/16x128b \\
\rowcolor{blue!5} AVX & x86/x86-64 & Intel & 2011 & 16x256b \\
\rowcolor{gray!10} XOP & x86/x86-64 & AMD & 2011 & 8/16x128b \\
\rowcolor{blue!5} AVX2 & x86/x86-64 & Intel & 2013 & 16x256b \\
\rowcolor{gray!10} AVX-512 & x86/x86-64 & Intel & 2016 & 32x512b \\
\hline
\end{tabular}
\caption{\ac{simd} extensions to general purpose processors \cite{simd_ref5}}
\label{tab:simd_ext}
\end{table}

\noindent Although \ac{simd} is quite fit for accelerating programs, there are still some limitations for using \ac{simd} in real-world applications.
\begin{itemize}
    \item Not all programs can be accelerated by \ac{simd} since not all programs can be vectorized easily. Some programs can be vectorized partially, but there would be a performance bottleneck due to the serial execution of non-vectorized functions.
    \item \ac{simd} are also limited for loading and storing data since all of the \ac{simd} \ac{isa}s have different vector widths.
    \item Usually, using \ac{simd} to accelerate the program needs a lot of human effort as the length of intrinsic code, or handwritten assembly code, is significantly longer than C or C++ code.
    \item Some \ac{simd} \ac{isa}s are more complicated to be used than others because of, for example, the restrictions on data alignment, not all processors supporting all the \ac{simd} \ac{isa}s. Investigating the influence of a new \ac{simd} for a program still needs a lot of human effort.
\end{itemize}

\subsection{Intel Streaming SIMD Extensions 4}
\label{sec:sse4}
Intel \ac{sse4} is a \ac{simd} \ac{cpu} instruction set which was announced at the Fall 2006 Intel Developer Forum fully compatible with software written for previous generations of Intel 64 and IA-32 architecture microprocessors. All existing software would continue to run correctly without modification on microprocessors that incorporate \ac{sse4}, as well as in the presence of existing and new applications that incorporate \ac{sse4}. Intel \ac{sse4} consists of 54 instructions. In some Intel documentation, a subset consisting of 47 instructions is referred to as \ac{sse4}.1. Additionally, \ac{sse4}.2, a second subset consisting of the seven remaining instructions.

\subsection{Intel Advanced Vector Extensions 2}
\label{sec:avx2}
\ac{avx2} expands most of the already introduced integer commands in \ac{sse4} to 256 bits and introduces \ac{fma} operations. Intel first supported them with the Haswell processor, which was introduced in 2013. \ac{avx2} uses 16 YMM registers to perform \ac{simd}. Each YMM register can hold and do simultaneous operations on eight 32-bit single-precision floating point numbers or four 64-bit double-precision floating point numbers. The \ac{sse} instructions are still utilized to operate on the lower 128 bits of the YMM registers. \ac{avx2} introduces a three-operand \ac{simd} instruction format, where the destination register is distinct from the two source operands.

\section{Related Work}
\label{sec:lit_survey_simd}
\begin{figure}[!ht]
    \centering
    \includegraphics[width=0.90\textwidth]{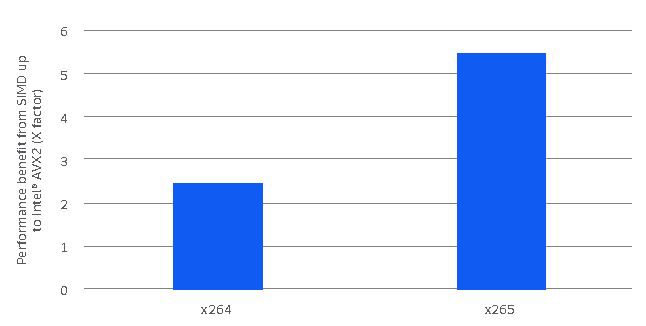}
    \caption{Performance benefit for x264 and x265 from Intel® Advanced Vector Extensions 2 for 1080p encoding with main profile using an Intel® Core™ i7-4500U Processor \cite{simd_ref6}}
    \label{fig:simd_x264_x265}
\end{figure}

The trend of optimizing programs by \ac{simd} extensions is quite common recently to leverage the existing \ac{cpu} architectures. For Intel \ac{cpu} architectures, different generations of \ac{simd} extensions such as \ac{sse2}, \ac{ssse3}, \ac{sse4}.1 and \ac{avx2} have been used to accelerate different video codecs such as \ac{mpeg}-2, \ac{mpeg}-4 Part 2. Recently H.264/AVC and \ac{hevc} have been optimized by the \ac{simd} instructions. A detailed summary of works reporting \ac{simd} optimization for codecs before H.264/\ac{avc} can be found in \cite{simd_ref1}. H.264/\ac{avc} also benefits greatly from \ac{simd} accelerating, including luma and chroma interpolation filters, inverse transform, and deblocking filter acceleration. For instance, Zhou et al. \cite{simd_ref2}, and Chen et al. \cite{simd_ref3} have reported that accelerating H.264/\ac{avc} by \ac{sse2} can speedup the application ranging from 2.0 to 4.0. One of my previous works was accelerating x265 \ac{hevc} encoder by AVX512 \ac{simd} extensions targeting Intel Core i9, and Intel® Xeon® Scalable Processors \cite{simd_ref6}. Figure \ref{fig:simd_x264_x265} shows the measure of encoder speedup in x264 and x265 encoders using Intel® \ac{avx2} \ac{simd} optimization. This motivates the x86 optimization of a more complex Arowana XVC encoder for better performance benefits.

%% file: kth_chapter4_analysis_methods.tex
\chapter{Analysis Sharing Methods in x265}
\label{sec:analysis_share_x265_method}
This chapter gives a brief insight into the methods used in this work for multi-resolution multi-rate analysis sharing in x265. Section \ref{sec:testing_method} illustrates how the various approaches considered in this work are tested. Section \ref{sec:intra_res_share_x265} explains the methods used in intra-resolution analysis sharing where Section \ref{sec:intra_res_stateofart}  details the state of the art method while Section \ref{sec:intra_analysis_median} describes the proposed method. Section \ref{sec:inter_res_share_x265} shows the details of the method used in Inter resolution analysis sharing.

\section{Testing method}
\label{sec:testing_method}
The evaluation of the methods used was performed on different types of sequences with varying content. In the report, I mention the results obtained using well-known
short sequences mentioned in Table \ref{tab:test_videos}.
\begin{table}[!htbp]
\centering
\begin{tabular}{ |p{3.8cm}| p{2.8cm}| p{1.8cm}| p{2.4cm}| p{1.8cm}|}
\hline
\multicolumn{5}{|c|}{Test Videos}\\
\hline
Video & Resolution & FPS & Colour Space & Frames\\
\hline
\rowcolor{gray!10} BasketBallDrive & 960x540 \newline 1920x1080 \newline 3840x2160 & 50 & YUV420p & 500\\
\rowcolor{blue!5} CrowdRun & 960x540 \newline 1920x1080 \newline 3840x2160 & 50 & YUV420p & 500\\
\rowcolor{gray!10} DucksTakeOff & 960x540 \newline 1920x1080 \newline 3840x2160 & 50 & YUV420p & 500\\
\hline
\end{tabular}
\caption{Videos used to validate the methods used}
\label{tab:test_videos}
\end{table}

\noindent In this work, the evaluations are focused on the medium preset of x265 \cite{x265_cli_ref}, which is the default quality preset in FFmpeg \cite{ff_ref} and presenting an optimum trade-off between the encoding speed and the compression efficiency.

\noindent For the testing of intra-resolution analysis sharing as mentioned in Section \ref{sec:intra_res_share_x265}, the test sequences are encoded for five different \ac{qp} representations for each resolution (540p and 1080p) as mentioned in Table \ref{tab:test_cqp_ladder}. Similarly, for the testing of inter-resolution analysis sharing as mentioned in Section \ref{sec:inter_res_share_x265}, the test sequences are encoded for five different \ac{qp} representations.  When the right approaches are determined for intra-resolution and inter-resolution analysis sharing, the \ac{cvbr} bitrates are used as mentioned in Table \ref{tab:test_bitrate_ladder} to investigate the effect of the approach in real-time streaming applications (shown in Section \ref{sec:Results}). The motivation to select five different \ac{qp} or \ac{cvbr} representations is to calculate the \ac{bdrate}, and \ac{bdpsnr} metrics \cite{bd_rate_ref} which gives us the proper understanding of the influence of the encoding approaches in terms of the compression efficiency. The results for 2160p encodes are calculated only for \ac{cvbr} representations.\\

\begin{table}[!htbp]
\centering
\begin{tabular}{ |p{3.2cm}|p{1.8cm}|p{1.8cm}|p{1.8cm}|p{1.8cm}|p{1.8cm}|}
\hline
\multicolumn{6}{|c|}{\ac{cqp} representations}\\
\hline
Resolution & \multicolumn{5}{c|}{\ac{qp}}\\
\hline
\rowcolor{gray!10} 960x540 & 22 & 26 & 30 & 34 & 38\\
\rowcolor{blue!5} 1920x1080 & 22 & 26 & 30 & 34 & 38\\
\rowcolor{gray!10} 3840x2160 & 22 & 26 & 30 & 34 & 38\\
\hline
\end{tabular}
\caption{\ac{qp} considered for \ac{cqp} mode test for various resolutions}
\label{tab:test_cqp_ladder}
\end{table}

\begin{table}[!htbp]
\centering
\begin{tabular}{ |p{3.2cm}|p{1.8cm}|p{1.8cm}|p{1.8cm}|p{1.8cm}|p{1.8cm}|}
\hline
\multicolumn{6}{|c|}{Bitrate representations}\\
\hline
Resolution & \multicolumn{5}{c|}{Bitrate (in kbps)}\\
\hline
\rowcolor{gray!10} 960x540 & 1000 & 1750 & 2500 & 3000 & 3500\\
\rowcolor{blue!5} 1920x1080 & 3500 & 4500 & 5500 & 7500 & 9000\\
\rowcolor{gray!10} 3840x2160 & 11000 & 13000 & 15000 & 17000 & 19000\\
\hline
\end{tabular}
\caption{\ac{cvbr} bitrates considered for various resolutions}
\label{tab:test_bitrate_ladder}
\end{table}

\noindent Bjøntegaard Delta (\ac{bdrate} and \ac{bdpsnr}) measurement \cite{bd_rate_ref} was used for evaluating the compression efficiency impact of the encoding approaches. \ac{bdrate} calculates the percentage difference in bitrate between \ac{rd} curves for the same quality. \ac{bdpsnr} calculates the average quality difference between \ac{rd} curves for the same bitrate. A negative \ac{bdpsnr} indicates a drop in coding efficiency of the proposed approach versus a reference approach, while a positive \ac{bdpsnr} represents a coding gain. Similarly, a positive \ac{bdrate} indicates a drop in coding efficiency of the proposed approach versus a reference approach, while a negative \ac{bdrate} represents a coding gain.\\

\noindent All performance and quality data for this work are collected on a system with Intel® Core (TM) i7-7500U \ac{cpu} running on Windows 10. Table \ref{tab:test_sytem_specs} presents the test system attributes.

\begin{table}[!htbp]
\centering
\begin{tabular}{ |p{3.2cm}|p{10.8cm}|}
\hline
\multicolumn{2}{|c|}{Test System details}\\
\hline
System Attribute & Value \\
\hline
\rowcolor{gray!10} OS Name & Windows 10 Home \\
\rowcolor{blue!5} Version & 10.0 Build 18363 \\
\rowcolor{gray!10} System Model & Lenovo 80YL \\
\rowcolor{blue!5} System Type & x64 based PC \\
\rowcolor{gray!10} Processor & Intel Core (TM) i7-7500U CPU @2.70Ghz (4 CPUs), ~2.9GHz\\
\rowcolor{blue!5} Memory Type & DDR4\\
\rowcolor{gray!10} Memory channel & 1 \\
\rowcolor{blue!5} Memory Size & 16384MB\\
\hline
\end{tabular}
\caption{System configurations for Testing}
\label{tab:test_sytem_specs}
\end{table}

\section{Intra resolution analysis sharing in x265}
\label{sec:intra_res_share_x265}
In this experiment, a multi-quality encoding system that encodes each video segment at multiple quality levels is modeled and studied. Section \ref{sec:intra_res_stateofart}  details the state-of-the-art method where the highest quality representation is the master/base representation. Section \ref{sec:intra_analysis_median} describes the method proposed to improve the state-of-the-art method.

\subsection{State of the art method}
\label{sec:intra_res_stateofart}
The highest quality representation is first encoded. The encoder analysis information is shared with the subsequent encodings with different \ac{qp}. This is the same as the setup used in the approaches mentioned in \cite{reuse_pap_5}, \cite{reuse_pap_3}, and \cite{reuse_pap_2}. Figure \ref{fig:intra_analysis_diag} illustrates the flowchart of the analysis-sharing scheme. It is also imperative that in an \ac{abr} system, the bitrate tiers are as close together as possible to avoid noticeable fluctuations in quality while switching across tiers. Here we have five \ac{qp} representations for 540p and 1080p encodings, respectively. For this model, the \ac{bdrate}, \ac{bdpsnr} \cite{bd_rate_ref}, and encoder speedup is evaluated for the overall encoding process.  

\tikzstyle{process2} = [rectangle, minimum width=2.5cm, minimum height=0.75cm, text centered, text width=2.5cm, draw=black, fill=orange!30]

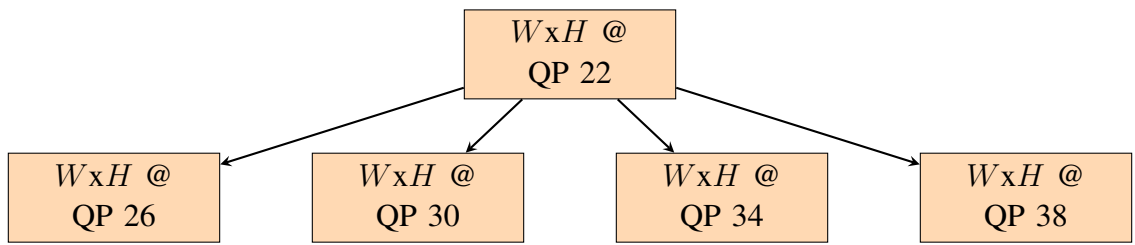
\begin{figure}[!htbp]
\begin{center}
\begin{tikzpicture}[node distance=1.0cm]
\node (i1) [process2] {$W$x$H$ @ QP 22};
\node (i2) [process2, below of=i1,xshift=-6cm, yshift = -0.9cm] {$W$x$H$ @ QP 26};
\node (i3) [process2, below of=i1,xshift=-2cm, yshift = -0.9cm] {$W$x$H$ @ QP 30};
\node (i4) [process2, below of=i1,xshift=2cm, yshift = -0.9cm] {$W$x$H$ @ QP 34};
\node (i5) [process2, below of=i1,xshift=6cm, yshift = -0.9cm] {$W$x$H$ @ QP 38};
\draw [arrow] (i1) -- (i2);
\draw [arrow] (i1) -- (i3);
\draw [arrow] (i1) -- (i4);
\draw [arrow] (i1) -- (i5);
\end{tikzpicture}
\end{center}
\caption{Intra resolution analysis sharing across $W$x$H$ resolution encode for \ac{cqp} representation with lowest \ac{qp} as master representation}
\label{fig:intra_analysis_diag}
\end{figure}

\noindent The encoding scheme is evaluated for all analysis reuse levels defined in x265 as shown in \ref{sec:x265_analysis_modes}. From Table \ref{tab:intra_resolution_short_result}, we see that x265 analysis reuse level 10 gives an average speedup of approximately 53 \% with an average \ac{bdrate} of 12 \% for 540p videos. Most of the state-of-the-art methods employ this reuse level. We also observe that analysis reuse level 10 gives a similar speedup for 1080p videos but with an increase in the \ac{bdrate}. But since we have a wide range of \ac{qp}s tested, there is more variance in the CU depth and mode decisions in different representations; hence, the rise in \ac{bdrate} is expected.

\begin{table}[!htbp]
\centering
\begin{tabular}{| p{2.0cm} |p{2.5cm}| p{2.0cm}| p{2.0cm} |p{2.0cm} |}
\hline
\multicolumn{5}{|c|}{Test for \ac{cqp} representations}\\
\hline
Resolution & Analysis level &	$\Delta$T & \ac{bdrate} & \ac{bdpsnr}\\
\hline
\rowcolor{blue!5} 960x540 & 4 &	 15.87 \%  &	\textbf{1.65 \%} & \textbf{-0.07 dB } \\
\rowcolor{gray!10} 960x540 & 6 & 17.23 \%   &  \textbf{1.65 \%} & -0.07 dB \\
\rowcolor{blue!5} 960x540 & 10 & 53.06 \% &	\textbf{12.27\%} & \textbf{-0.46 dB } \\
\hline
\rowcolor{gray!10} 1920x1080 & 4 &	 11.57 \%  &	\textbf{2.03 \%} & \textbf{-0.06 dB } \\
\rowcolor{blue!5} 1920x1080 & 6 &	12.83 \%   &  \textbf{2.03 \%} & \textbf{-0.06 dB } \\
\rowcolor{gray!10} 1920x1080 & 10 & 49.99 \% & \textbf{21.69 \%} & \textbf{-0.47 dB } \\
\hline
\end{tabular}
\caption{Measure of speedup and compression efficiency for intra-resolution analysis sharing with lowest \ac{qp} (highest quality) representation as master encode}
\label{tab:intra_resolution_short_result}
\end{table}

\subsection{Proposed method}
\label{sec:intra_analysis_median}
In the model described in the previous section, in compliance with most of the literature, the highest quality encode is the master representation. There are two key problems to address here:
\begin{itemize}
    \item In parallel encoding systems, by this approach, the overall performance is limited by the time taken for the master encoding. Hence, using the highest quality encode as the master representation causes performance bottlenecks for multi-quality systems.
    \item The CU depth and mode decisions from the highest quality representation are sub-optimal as we traverse towards lower quality representations. 
\end{itemize}

\begin{figure}[!htbp]
\begin{center}
\begin{tikzpicture}[node distance=1.0cm]
\node (i1) [process2] {$W$x$H$ @ QP 30};
\node (i2) [process2, below of=i1,xshift=-6cm, yshift = -0.9cm] {$W$x$H$ @ QP 22};
\node (i3) [process2, below of=i1,xshift=-2cm, yshift = -0.9cm] {$W$x$H$ @ QP 26};
\node (i4) [process2, below of=i1,xshift=2cm, yshift = -0.9cm] {$W$x$H$ @ QP 34};
\node (i5) [process2, below of=i1,xshift=6cm, yshift = -0.9cm] {$W$x$H$ @ QP 38};
\draw [arrow] (i1) -- (i2);
\draw [arrow] (i1) -- (i3);
\draw [arrow] (i1) -- (i4);
\draw [arrow] (i1) -- (i5);
\end{tikzpicture}
\end{center}
\caption{Intra resolution analysis sharing across $W$x$H$ resolution encode for \ac{cqp} representation with median \ac{qp} as master representation}
\label{fig:intra_analysis_median_diag}
\end{figure}
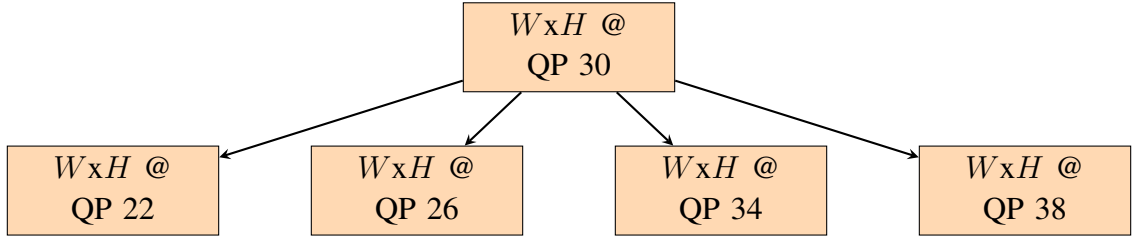

\noindent To mitigate the above-mentioned issues, the experiment is conducted using the median quality representation as the master representation as shown in Figure \ref{fig:intra_analysis_median_diag}.

\noindent For the five \ac{qp} representations in the test, \ac{qp} 30 is the median quality representation. Hence, \ac{qp} 30 representation is encoded first. The encoder analysis information is shared with the subsequent encodings- \ac{qp} 22, 26, 34, and 38. Typically, \ac{qp} 22 standalone representation is the slowest among all stand-alone encodings, resulting in a performance bottleneck for parallel encoding systems. But here, \ac{qp} 22 encodings are expected to be paced up by using the encoder analysis information of the \ac{qp} 30 representation.\\

\noindent Table \ref{tab:intra_resolution_median_short_result} shows the speedup and \ac{bdrate} for 540p and 1080p intra-resolution analysis sharing with the proposed method. Table \ref{tab:bdrate_intra_resolution_result} compares the \ac{bdrate} and \ac{bdpsnr} for the state-of-the-art method and proposed method. There is a significant reduction in the \ac{bdrate} for the proposed method. Hence, the median quality representation as the master representation approach is better than the existing state-of-the-art highest quality representation as the master representation approach.

\begin{table}[!htbp]
\centering
\begin{tabular}{| p{2.0cm} |p{2.5cm}| p{2.0cm}| p{2.0cm} |p{2.0cm} |}
\hline
\multicolumn{5}{|c|}{Test for \ac{cqp} representations}\\
\hline
Resolution & Analysis level &	$\Delta$T & \ac{bdrate} & \ac{bdpsnr}\\
\hline
\rowcolor{gray!10} 960x540 & 4 &	 16.94 \%   & 0.80 \%  & -0.03 dB \\
\rowcolor{blue!5} 960x540 & 6 &	18.77 \%   &  0.80 \%  &	-0.03 dB \\
\rowcolor{gray!10} 960x540 & 10 & 65.60 \% &	6.45 \%   &	-0.26 dB \\
\hline
\rowcolor{blue!5} 1920x1080 & 4 & 13.91 \%  &	1.61 \% &  -0.05 dB \\
\rowcolor{gray!10} 1920x1080 & 6 &	15.07 \%   &  1.61 \% &  -0.05 dB \\
\rowcolor{blue!5} 1920x1080 & 10 & 63.91 \%  &	8.59 \% &	-0.25 dB \\
\hline
\end{tabular}
\caption{Measure of speedup and compression efficiency for intra-resolution analysis sharing with median \ac{qp} representation as master encode}
\label{tab:intra_resolution_median_short_result}
\end{table}

\begin{figure}[!ht]
    \centering
    \includegraphics[width=0.90\textwidth]{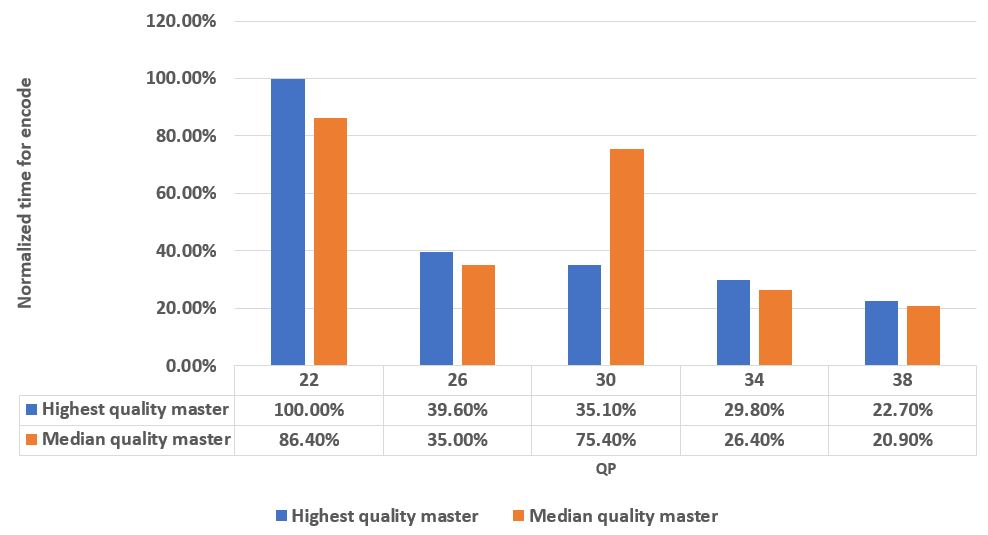}
    \caption{Average encode time for all representations in intra-resolution analysis sharing for 1080p videos}
    \label{fig:qp_intra_analysis}
\end{figure}

\noindent As mentioned earlier, the performance bottleneck of the state-of-the-art method was that the overall turnaround time in parallel encoding was bound by the speed of \ac{qp} 22 encode. So, a detailed speedup analysis for every resolution tier representation is needed. Figure \ref{fig:qp_intra_analysis} shows the speedup of each \ac{cqp} representation for 1080p encode with x265 analysis share mode 10 for the state-of-the-art method and the proposed method. All speeds are normalized to the \ac{qp} 22 stand-alone speed. In parallel encoding, the time required to encode these five representations is the time taken by the \ac{qp} 22 encode for the state-of-the-art method. For the proposed method, the time taken for \ac{qp} 22 representation is the total time required for encoding the five representations, but it is observed that that time taken is decreased to a factor of 86\%. It means that a factor of 14 \% paces up the parallel encoding. 

\begin{table}[!htbp]
\centering
\begin{tabular}{|p{2.4cm} | p{2.4cm} |p{1.8cm}| p{1.8cm}| p{1.8cm} |p{1.8cm} |}
\hline
& & \multicolumn{2}{c|}{Highest quality master} & \multicolumn{2}{c|}{Median quality master}\\
\hline
Resolution & Analysis level & \ac{bdrate} & \ac{bdpsnr} & \ac{bdrate} & \ac{bdpsnr}\\
\hline
\rowcolor{blue!5}  960x540 & 4  &	1.65 \% & -0.07 & 0.80 \% &	-0.03 dB \\
\rowcolor{gray!10} 960x540 & 6  &	1.65 \% & -0.07 & 0.80 \% &	-0.03 dB \\
\rowcolor{blue!5}  960x540 & 10 & 12.27 \% &  -0.46 & 6.45 \% &	-0.26 dB \\
\rowcolor{gray!10} 1920x1080 &  4 &	2.03 \% & -0.06 & 1.61 \% &	-0.05 dB \\
\rowcolor{blue!5}  1920x1080 &  6 &	2.03 \% & -0.06 & 1.61 \% &	-0.05 dB \\
\rowcolor{gray!10} 1920x1080 & 10 &	21.69 \% &-0.47 & 8.59 \% &	-0.25 dB \\
\hline
\end{tabular}
\caption{Average improvement of \ac{bdrate} for intra-resolution analysis sharing with median quality representation as master encode}
\label{tab:bdrate_intra_resolution_result}
\end{table}

\section{Inter-resolution analysis sharing in x265}
\label{sec:inter_res_share_x265}
In this experiment, a system is modeled which encodes each video segment at the highest quality in the lower resolution, and then reusing the frame-level metadata of the analysis information, the representations in the current resolution tier are encoded. This setup is schematically represented in Figure \ref{fig:inter_analysis_diag}. For this model, the \ac{bdrate}, \ac{bdpsnr}, and overall encoder speedup are evaluated for the current resolution tier encoding process.\\

\tikzstyle{process4} = [rectangle, minimum width=2.1cm, minimum height=0.75cm, text centered, text width=2.0cm, draw=black, fill=orange!30]

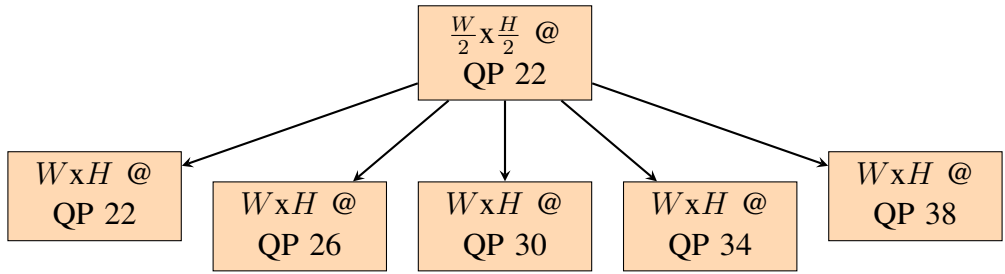
\begin{figure}[!htbp]
\begin{center}
\begin{tikzpicture}[node distance=1.0cm]
\node (i1) [process4] {$\frac{W}{2}$x$\frac{H}{2}$ @ QP 22};
\node (i2) [process4, below of=i1,xshift=-5.4cm, yshift = -0.9cm] {$W$x$H$ @ QP 22};
\node (i3) [process4, below of=i1,xshift=-2.7cm, yshift = -1.3cm] {$W$x$H$ @ QP 26};
\node (i4) [process4, below of=i1,yshift = -1.3cm] {$W$x$H$ @ QP 30};
\node (i5) [process4, below of=i1,xshift=2.7cm, yshift = -1.3cm] {$W$x$H$ @ QP 34};
\node (i6) [process4, below of=i1,xshift=5.4cm, yshift = -0.9cm] {$W$x$H$ @ QP 38};
\draw [arrow] (i1) -- (i2);
\draw [arrow] (i1) -- (i3);
\draw [arrow] (i1) -- (i4);
\draw [arrow] (i1) -- (i5);
\draw [arrow] (i1) -- (i6);
\end{tikzpicture}
\end{center}
\caption{Inter resolution analysis sharing for \ac{cqp} representation}
\label{fig:inter_analysis_diag}
\end{figure}

\begin{table}[!htbp]
\centering
\begin{tabular}{| p{2.0cm} | p{2.0cm} | p{1.5cm}| p{1.5cm} |p{1.8cm} |}
\hline
\multicolumn{5}{|c|}{Test for \ac{cqp} representations}\\
\hline
Resolution & Refinement & $\Delta$T & \ac{bdrate} & \ac{bdpsnr}\\
\hline
1920x1080 & no & \textbf{72.04 \% }  &	\textbf{20.08 \%}  & \textbf{-0.62 dB}\\
1920x1080 & yes & \textbf{28.99 \% }  &	\textbf{1.61 \%}  & \textbf{-0.07 dB}\\
\hline
\end{tabular}
\caption{Measure of speedup and \ac{bdrate} for 1080p inter-resolution analysis sharing with 540p representation as the master encode and analysis reuse level 10}
\label{tab:inter_resolution_1080p_result}
\end{table}

\noindent From Table \ref{tab:inter_resolution_1080p_result}, it is observed that, by reusing the analysis information from the highest quality representation of the 540p encode without analysis refinement, the 1080p resolution tier was encoded with an overall speedup of 72.04 \% with \ac{bdrate} of 20.08 \% and \ac{bdpsnr} of -0.62 dB. It is also observed that, by reusing the analysis information from the highest quality representation of the 540p encode with analysis refinement, the 1080p resolution tier was encoded with an overall speedup of 28.99 \% with \ac{bdrate} of 1.61\% and \ac{bdpsnr} of -0.07 dB. Hence, it is observed that refining the analysis across the resolutions is an excellent strategy to have a decent speedup and negligible \ac{bdrate}.

%% file: kth_chapter5_simd_methods.tex
\chapter{Intel SIMD optimization methods for XVC Encoding}
\label{sec:xvc_simd_method}
This chapter presents the exploratory work done to understand the impact of \ac{simd} parallelism in media processing. Section \ref{sec:simd_impl_methods} shows various methods used for \ac{simd} implementation and motivates the choice of the hand-written assembly method. Section \ref{sec:nasm} explains about NASM assembler used for compiling hand-written assembly code in this project. Section \ref{sec:acc_xvc_simd} shows the \ac{cpu} cycle count improvements for kernels accelerated with Intel \ac{sse4} and \ac{avx2} \ac{simd}. Section \ref{sec:sel_kernel} mentions how the functions to be optimized are chosen, Section \ref{sec:cost_func} and Section \ref{sec:simd_cost_func} describing the implementation of encoder cost functions and the resulting \ac{cpu} cycle count improvements. Section \ref{sec:cycl_improve_method} explains the method used to evaluate the \ac{cpu} cycle count improvements for each optimized function.

\section{SIMD implementation methods}
\label{sec:simd_impl_methods}
There are three primary methods of utilizing \ac{simd} instructions on Intel processors:
\begin{itemize}
    \item writing low-level assembly code
    \item use of compiler-supported intrinsic functions
    \item compiler auto-vectorization
\end{itemize}
Generally, writing low-level assembly code that uses \ac{simd} instructions and available registers is considered the best approach for achieving high performance. However, this method is cumbersome and error-prone. Compiler-supported intrinsic functions provide a higher level of abstraction, with an almost one-to-one mapping to assembly instructions, but without the need to deal with register allocations, instruction scheduling, type checking, and call stack maintenance. In this approach, intrinsic functions are expanded inline, eliminating function call overhead. It provides the same benefits as inline assembly, improved code readability, and fewer errors \cite{asm_ref}. The penalty is that the overall performance boost depends on the compiler's ability to optimize across multiple intrinsic function calls. Compiler auto-vectorization leaves everything to the compiler, relying on it to locate and automatically vectorize suitable loop structures. Therefore, the compiler's quality and the programmer's ability to write code that aids auto-vectorization becomes essential. \cite{asm_ref} showed that state-of-the-art compilers were able to vectorize only 18-30 \% of real application codes. In particular, it was noted that compilers did not perform some critical code transformations necessary for facilitating auto-vectorization. Non-unit stride memory access, data alignment, and data dependency transformations were found to be pertinent issues with compiler auto-vectorization. We choose the low-level assembly code method to perform \ac{simd} operations in our work.

\section{NASM Assembler support}
\label{sec:nasm}
The assembler support must be added to compile the hand-written assembly code to the video encoding solution. In this project, the support of the Netwide Assembler (NASM) is added to the Arowana XVC encoder solution \cite{nasm_ref}. NASM is an assembler and disassembler for the Intel x86 architecture, which can be used to write 16-bit, 32-bit (IA-32), and 64-bit (x86-64) programs. It is open-source software released under the terms of a simplified (2-clause) BSD license. The required changes are made to the solution compiling method so that NASM generates the compiled object files of the assembly code, and they get linked to the encoder solution.

\section{Accelerating Arowana XVC Encoding with Intel SIMD}
\label{sec:acc_xvc_simd}
\begin{figure}[!ht]
    \centering
    \includegraphics[width=0.91\textwidth]{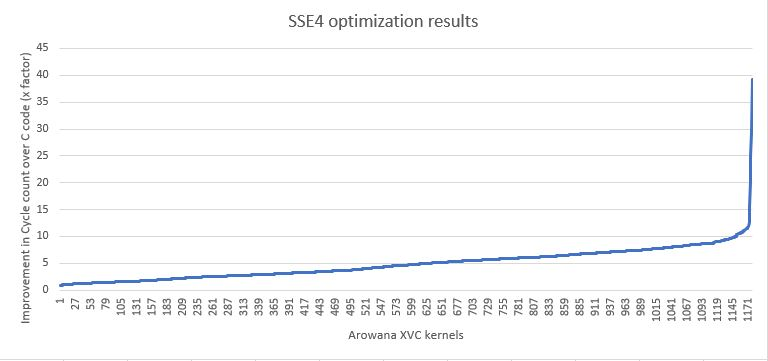}
    \caption{Cycle-count gains of the Intel® Streaming SIMD Extensions 4 kernels over the corresponding C++ kernels}
    \label{fig:sse4_kernels_graph}
\end{figure}

Figure \ref{fig:sse4_kernels_graph} shows the cycle-count improvements for each of the kernels that were accelerated with \ac{sse4} Intel \ac{simd}. The kernels are sorted in increasing order of their cycle count gains over the corresponding C++ implementation. The detailed performance improvement for each optimized kernel gains over its corresponding C++ implementation is shown in Appendix \ref{sec:sse4_test_res}.\\

\begin{figure}[!ht]
    \centering
    \includegraphics[width=0.91\textwidth]{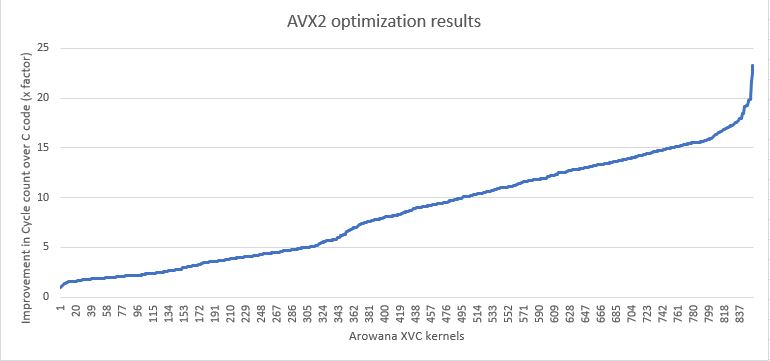}
    \caption{Cycle-count gains of the Intel® Advanced Vector Extensions 2 kernels over the corresponding C++ kernels}
    \label{fig:avx2_kernels_graph}
\end{figure}

\noindent Figure \ref{fig:avx2_kernels_graph} shows the cycle-count improvements for each of the kernels that were accelerated with \ac{avx2} Intel \ac{simd}. The detailed performance improvement for each optimized kernel gains over its corresponding C++ implementation is shown in Appendix \ref{sec:avx2_test_res}.

\noindent The results from cycle counts indicate that at the kernel level, using Intel \ac{simd} to accelerate XVC is advantageous.

\section{Selecting the kernels to Accelerate}
\label{sec:sel_kernel}
Over 1100+ kernels were selected from the core compute of XVC to optimize with Intel \ac{simd} instructions for the main10 profile. These kernels were chosen based on their resource requirements. Some kernels may require frequent memory access, like different block-copy and block-fill kernels, while others may involve intense computation, like DCT, iDCT, and quantization kernels. There is also a third class of kernels that involves a combination of both in varying proportions. A complete list of the kernels optimized with Intel \ac{simd} instructions are listed in Section \ref{sec:sse4_test_res} and Section \ref{sec:avx2_test_res}.

\subsection{Cost functions}
\label{sec:cost_func}
SAD (sum of absolute difference), SATD (sum of absolute transform difference), and DCT are some of the most complex and frequently called functions in the Arowana XVC encoder. In video coding, many cost functions are used to evaluate and choose the best coding mode and its associated parameters. SAD and SATD are the two main cost functions to find integer and quarter-pel motion vectors in motion estimation. SAD takes around 11 \% to 13 \% of the total \ac{cpu} cycles in the XVC encoder, and SATD takes 14 \% to 15 \%. These two cost functions are defined by:
\begin{equation}
    SAD = \Sigma_{i,j}^{I,J} \mid O(i, j) - P(i, j) \mid
\end{equation}
\begin{equation}
    SATD = \frac{\Sigma_{i,j}^{I,J} \mid H(i, j) \mid}{2}
\end{equation}
where i and j are the pixel indices, and their ranges are determined by block size. O(i,j) and P(i,j) is the original and predicted pixel values, respectively. H(i,j) is the Hadamard transform of the prediction error, O(i,j) - P(i,j) \cite{ref:sad_asm}. Because only addition and subtraction operations are involved in the cost functions, SATD can yield an accurate cost in the transform domain with relatively small complexity compared to DCT. Since both apply the same operations on multiple data, vector instructions are quite useful for reducing the required clock cycles.

\subsection{SIMD implementation of cost functions}
\label{sec:simd_cost_func}
\subsubsection{SAD function}
The C++ code for the SAD function is shown below:
\begin{lstlisting}[style=CStyle]
template<int width, int height, typename SampleT1>
int ComputeSad_c(const SampleT1 *sample1, ptrdiff_t stride1,
                 const Sample *sample2, ptrdiff_t stride2) {
  int sum = 0;
  for (int y = 0; y < height; y++) {
    for (int x = 0; x < width; x++) {
      int diff = sample1[x] - sample2[x];
      sum += std::abs(diff);
    }
    sample1 += stride1;
    sample2 += stride2;
  }
  return sum;
}
\end{lstlisting}

\noindent The corresponding x86 handwritten assembly for block sizes of width 16 is shown below:
\begin{lstlisting}[style=CStyle]
%macro SAD_16x2_HBD_AVX2 1
    pxor    m0, m0
    mova    m6, [pw_1]
    shl     r3d, 1
    shl     r1d, 1
%rep %1/2
    movu    m1, [r2]
    movu    m2, [r2 + r3]
    psubw   m1, [r0]
    psubw   m2, [r0 + r1]
    pabsw   m1, m1
    pabsw   m2, m2
    paddw   m1, m2
    pmaddwd m1, m6
    paddd   m0, m1
    lea     r0, [r0 + 2 * r1]
    lea     r2, [r2 + 2 * r3]
%endrep
    HADDD   m0, m1
    movd    eax, xm0
%endmacro

INIT_YMM avx2
cglobal sad_16x2, 4, 4, 7
    SAD_16x2_HBD_AVX2 2
    RET
cglobal sad_16x4, 4, 4, 7
    SAD_16x2_HBD_AVX2 4
    RET
cglobal sad_16x8, 4, 4, 7
    SAD_16x2_HBD_AVX2 8
    RET
cglobal sad_16x16, 4, 4, 7
    SAD_16x2_HBD_AVX2 16
    RET
cglobal sad_16x32, 4, 4, 7
    SAD_16x2_HBD_AVX2 32
    RET
cglobal sad_16x64, 4, 4, 7
    SAD_16x2_HBD_AVX2 64
    RET
\end{lstlisting}

\noindent Table \ref{tab:sad_speedup} shows the performance gains from the optimized kernels of the SAD function over its C++ implementation. 
\begin{table}[!htbp]
\centering
\begin{tabular}{ |p{2.6cm}| p{1.5cm}| p{2.6cm}| p{1.5cm}|p{2.6cm}| p{1.5cm}|}
\hline
\multicolumn{6}{|c|}{Speedup with \ac{sse4} and \ac{avx2} optimizations}\\
\hline
Kernel & Speedup & Kernel & Speedup & Kernel & Speedup\\
\hline
\rowcolor{gray!10} sad[4x4] &     3.01x &  sad[8x64] &     2.63x &   sad[32x8] &     4.95x\\
\rowcolor{blue!5}  sad[4x8] &     4.41x &  sad[16x2] &     3.39x &  sad[32x16]  &    4.99x \\
\rowcolor{gray!10} sad[4x16] &     3.67x &  sad[16x4] &     2.44x &   sad[32x32] &     5.80x\\  
\rowcolor{blue!5}  sad[4x32] &     5.54x &  sad[16x8] &     4.58x &   sad[32x64] &     5.54x\\ 
\rowcolor{gray!10} sad[8x4] &     2.90x &  sad[16x16] &    6.46x &   sad[64x8]  &    5.30x\\
\rowcolor{blue!5}  sad[8x8] &     2.40x &  sad[16x32] &    5.79x &  sad[64x16] &     5.37x\\
\rowcolor{gray!10} sad[8x16] &     5.72x &  sad[16x64] &    6.27x &  sad[64x32] &     5.27x\\
\rowcolor{blue!5}  sad[8x32] &     2.78x &  sad[32x4] &     4.47x &	  sad[64x64] &     5.14x\\
\hline
\end{tabular}
\caption{Speedups of SAD kernels with \ac{sse4} and \ac{avx2} optimizations}
\label{tab:sad_speedup}
\end{table}

\subsubsection{SATD function}
The C++ code for the SATD function is shown below:
\begin{lstlisting}[style=CStyle]
template<typename SampleT1, int w, int h>
// calculate satd in blocks of 8x4
int satd8(const SampleT1* pix1, intptr_t stride_pix1,
          const Sample* pix2, intptr_t stride_pix2) {
  int satd = 0;
  for (int row = 0; row < h; row += 4)
    for (int col = 0; col < w; col += 8)
      satd += satd_8x4(pix1 + row * stride_pix1 + col, stride_pix1,
        pix2 + row * stride_pix2 + col, stride_pix2);
  return satd;
}

#define HADAMARD4(d0, d1, d2, d3, s0, s1, s2, s3) { \
        uint64_t t0 = s0 + s1; \
        uint64_t t1 = s0 - s1; \
        uint64_t t2 = s2 + s3; \
        uint64_t t3 = s2 - s3; \
        d0 = t0 + t2; \
        d2 = t0 - t2; \
        d1 = t1 + t3; \
        d3 = t1 - t3; \
}

template<typename SampleT1>
static int satd_8x4(const SampleT1* pix1, intptr_t stride_pix1,
                    const Sample* pix2, intptr_t stride_pix2) {
  uint64_t tmp[4][4];
  uint64_t a0, a1, a2, a3, b0, b1;
  uint64_t sum = 0;

  for (int i = 0; i < 4; i++, pix1 += stride_pix1, pix2 += stride_pix2) {
    b0 = pix1[0] - pix2[0];
    b1 = pix1[4] - pix2[4];
    a0 = b0 + (static_cast<uint64_t>(b1) << BITS_PER_SUM);
    b0 = pix1[1] - pix2[1];
    b1 = pix1[5] - pix2[5];
    a1 = b0 + (static_cast<uint64_t>(b1) << BITS_PER_SUM);
    b0 = pix1[2] - pix2[2];
    b1 = pix1[6] - pix2[6];
    a2 = b0 + (static_cast<uint64_t>(b1) << BITS_PER_SUM);
    b0 = pix1[3] - pix2[3];
    b1 = pix1[7] - pix2[7];
    a3 = b0 + (static_cast<uint64_t>(b1) << BITS_PER_SUM);
    HADAMARD4(tmp[i][0],tmp[i][1],tmp[i][2],tmp[i][3],a0,a1,a2,a3);
  }
  for (int i = 0; i < 4; i++) {
    HADAMARD4(a0,a1,a2,a3,tmp[0][i],tmp[1][i],tmp[2][i],tmp[3][i]);
    sum += abs2(a0) + abs2(a1) + abs2(a2) + abs2(a3);
  }
  return (static_cast<uint32_t>(sum) + (sum >> BITS_PER_SUM)) >> 1;
}
\end{lstlisting}

\noindent The corresponding x86 handwritten assembly for block sizes of width 16 is shown below:
\begin{lstlisting}[style=CStyle]
cglobal calc_satd_16x4    ; function to compute satd cost for 16 columns, 4 rows
    ; rows 0-3
    movu            m0, [r0]
    movu            m4, [r2]
    psubw           m0, m4
    movu            m1, [r0 + r1]
    movu            m5, [r2 + r3]
    psubw           m1, m5
    movu            m2, [r0 + r1 * 2]
    movu            m4, [r2 + r3 * 2]
    psubw           m2, m4
    movu            m3, [r0 + r4]
    movu            m5, [r2 + r5]
    psubw           m3, m5
    lea             r0, [r0 + r1 * 4]
    lea             r2, [r2 + r3 * 4]
    paddw           m4, m0, m1
    psubw           m1, m0
    paddw           m0, m2, m3
    psubw           m3, m2
    punpckhwd       m2, m4, m1
    punpcklwd       m4, m1
    punpckhwd       m1, m0, m3
    punpcklwd       m0, m3
    paddw           m3, m4, m0
    psubw           m0, m4
    paddw           m4, m2, m1
    psubw           m1, m2
    punpckhdq       m2, m3, m0
    punpckldq       m3, m0
    paddw           m0, m3, m2
    psubw           m2, m3
    punpckhdq       m3, m4, m1
    punpckldq       m4, m1
    paddw           m1, m4, m3
    psubw           m3, m4
    punpckhqdq      m4, m0, m1
    punpcklqdq      m0, m1
    pabsw           m0, m0
    pabsw           m4, m4
    pmaxsw          m0, m0, m4
    punpckhqdq      m1, m2, m3
    punpcklqdq      m2, m3
    pabsw           m2, m2
    pabsw           m1, m1
    pmaxsw          m2, m1
    pxor            m7, m7
    mova            m1, m0
    punpcklwd       m1, m7
    paddd           m6, m1
    mova            m1, m0
    punpckhwd       m1, m7
    paddd           m6, m1
    pxor            m7, m7
    mova            m1, m2
    punpcklwd       m1, m7
    paddd           m6, m1
    mova            m1, m2
    punpckhwd       m1, m7
    paddd           m6, m1
    ret

%macro SATD_AVX2_END 0
    vextracti128    xm7, m6, 1
    paddd           xm6, xm7
    pxor            xm7, xm7
    movhlps         xm7, xm6
    paddd           xm6, xm7
    pshufd          xm7, xm6, 1
    paddd           xm6, xm7
    movd            eax, xm6
%endmacro

;static int satd_16x4(const SampleT1* pix1, intptr_t stride_pix1,
;                     const Sample* pix2, intptr_t stride_pix2,
;                     int offset)
%macro SATD_16x4_AVX2 1
cglobal satd_%1_16x4, 4,6,8
    shl             r1d, 1
    shl             r3d, 1
    lea             r4, [3 * r1]
    lea             r5, [3 * r3]
    pxor            m6, m6
    call            calc_satd_16x4
    SATD_AVX2_END
    RET
%endmacro
\end{lstlisting}

\begin{table}[!htbp]
\centering
\begin{tabular}{ |p{2.6cm}| p{1.5cm}| p{2.6cm}| p{1.5cm}|p{2.6cm}| p{1.5cm}|}
\hline
\multicolumn{6}{|c|}{Speedup with \ac{sse4} and \ac{avx2} optimizations}\\
\hline
Kernel & Speedup & Kernel & Speedup & Kernel & Speedup\\
\hline
\rowcolor{gray!10}   satd[8x4]  &    3.45x  &    satd[16x16] &     7.32x  &   satd[32x64] &     8.72x \\
\rowcolor{blue!5}   satd[8x8]  &    3.75x &	  satd[16x32] &     8.15x &      satd[64x8] &     8.43x	\\		   
\rowcolor{gray!10}   satd[8x16] &     4.21x &     satd[16x64] &     8.85x &    satd[64x16] &     9.03x\\				\rowcolor{blue!5}    satd[8x32] &     4.01x &      satd[32x4] &     5.61x &    satd[64x16] &     9.19x\\
\rowcolor{gray!10}   satd[8x64] &     4.40x &      satd[32x8] &     7.83x &    satd[64x32] &     9.81x\\
\rowcolor{blue!5}    satd[16x4] &     5.19x &     satd[32x16] &     7.75x &    satd[64x64] &     10.22x	\\
\rowcolor{gray!10}   satd[16x8] &     6.40x &     satd[32x32] &     8.56x & & \\
\hline
\end{tabular}
\caption{Speedups of SATD kernels with \ac{sse4} and \ac{avx2} optimizations}
\label{tab:satd_speedup}
\end{table}

\noindent Table \ref{tab:satd_speedup} shows the performance gains from the optimized kernels of the SATD function over its C++ implementation. 3x to 10x speedup is observed for various kernel sizes using \ac{sse4} and \ac{avx2} \ac{simd} extensions.
                      
\section{Method to evaluate cycle-count improvements}
\label{sec:cycl_improve_method}
When handwritten x86 \ac{simd} assembly code is implemented, two criteria need to be evaluated:
\begin{itemize}
    \item If the primitive implemented is correct, i.e. if the outputs are the same, the same set of inputs is fed into the C++ primitive and the optimized primitive.
    \item If the primitive implemented is correct, the speedup is obtained from the optimized primitive with respect to the C++ primitive.
\end{itemize}
In this case, a correctness and performance measurement tool is implemented in the Arowana XVC encoder for x86 assembly kernels. To evaluate the first criteria mentioned above, the tool accepts valid arguments for a given primitive, invokes the C++ primitive and corresponding assembly kernel, and compares both output buffers. It verifies all possible corner cases for the given input type using a randomly distributed set of values. Each assembly kernel is called 100 times and checked against its C++ primitive output to ensure correctness. For the second criterion, ie to measure performance improvement, the test bench measures the difference in the clock ticks (as reported by the rdtsc instruction) between the assembly kernel and the C++ kernel for 1,000 runs and reports the average between them.

%% file: kth_chapter6_results.tex
\chapter{Results and Discussion}
\label{sec:Results}
In this chapter, the methods described in the previous chapters are evaluated in real-time streaming scenarios. In the state-of-the-art applications in the streaming industry, \ac{cvbr} mode is used for video encoding. Hence, this section presents the results of the methods mentioned in Section \ref{sec:analysis_share_x265_method} using \ac{cvbr} mode in x265. The \ac{cvbr} bitrates used for testing for each resolution are shown in Table \ref{tab:test_bitrate_ladder}.  Section \ref{sec:intra_abr} shows the state-of-the-art intra-resolution analysis sharing method and discusses the results for the considered resolutions- 540p, 1080p, and 2160p. Section \ref{sec:novel_intra_abr} discusses the proposed intra-resolution analysis sharing method and its improvement over the traditional intra-resolution analysis sharing method. Section \ref{sec:inter_abr} explains the inter-resolution analysis sharing method, the need for analysis refinement, and the associated results. Section \ref{sec:stateofart_multi_res_rate} and Section \ref{sec:proposed_multi_res_rate} discuss state-of-the-art and the proposed multi-encoder structure with the scheme for encoder analysis sharing and the associated results, respectively. Section \ref{sec:xvc_simd_res} explains the improvements of the Arowana XVC encoding speeds utilizing the \ac{simd} extensions.

\section{State of the art Intra-resolution analysis sharing in x265}
\label{sec:intra_abr}

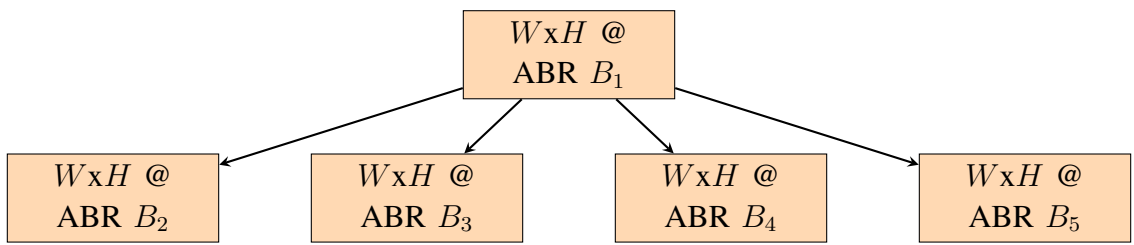
\begin{figure}[!htbp]
\begin{center}
\begin{tikzpicture}[node distance=1.0cm]
\node (i1) [process2] {$W$x$H$ @ ABR $B_{1}$};
\node (i2) [process2, below of=i1,xshift=-6cm, yshift = -0.9cm] {$W$x$H$ @ ABR $B_{2}$};
\node (i3) [process2, below of=i1,xshift=-2cm, yshift = -0.9cm] {$W$x$H$ @ ABR $B_{3}$};
\node (i4) [process2, below of=i1,xshift=2cm, yshift = -0.9cm] {$W$x$H$ @ ABR $B_{4}$};
\node (i5) [process2, below of=i1,xshift=6cm, yshift = -0.9cm] {$W$x$H$ @ ABR $B_{5}$};
\draw [arrow] (i1) -- (i2);
\draw [arrow] (i1) -- (i3);
\draw [arrow] (i1) -- (i4);
\draw [arrow] (i1) -- (i5);
\end{tikzpicture}
\end{center}
\caption{Intra resolution analysis sharing across $W$x$H$ resolution encode for \ac{abr} representation with the state of the art method ($B_{1}$ > $B_{2} > $ $B_{3}$ > $B_{4}$ > $B_{5}$)}
\label{fig:intra_analysis_abr_diag}
\end{figure}

In this model, the \ac{cvbr} $B_{1}$ encode instance is the master representation. The subsequent encodes uses the analysis information from the master representation. The results are evaluated for the test videos mentioned in Table \ref{tab:test_videos} and \ac{cvbr} bitrates shown in Table \ref{tab:test_bitrate_ladder}. Here, only x265 analysis reuse level 10 mentioned in Table \ref{tab:x265_analysis} is used since all recent literature use that analysis sharing mode.\\

\noindent It is observed that the results are very similar to the ones for \ac{cqp} representation. As shown in Table \ref{tab:intra_resolution_abr_result}, overall speedup for the five \ac{cvbr} representations for 540p videos is $54.00 \%$ with a \ac{bdrate} of $8.79 \%$ and \ac{bdpsnr} of $-0.36 dB$. For 1080p videos, the overall speedup is $54.62 \%$ with a \ac{bdrate} of $9.23 \%$ and \ac{bdpsnr} of $-0.27 dB$. For 2160p videos, the overall speedup is $58.00 \%$ with a \ac{bdrate} of $8.22 \%$ and \ac{bdpsnr} of $-0.25 dB$. 

\begin{table}[!htbp]
\centering
\begin{tabular}{| p{2.0cm} | p{2.0cm}| p{2.0cm} |p{2.0cm} |}
\hline
\multicolumn{4}{|c|}{Test for \ac{abr} representations}\\
\hline
Resolution & $\Delta$T & \ac{bdrate} & \ac{bdpsnr}\\
\hline
\rowcolor{gray!10} 960x540 &  54.00 \%   &	8.79 \% & -0.36 dB \\
\rowcolor{blue!5} 1920x1080 &  54.62 \%   &	9.23 \% & -0.27 dB \\
\rowcolor{gray!10} 3840x2160 &  58.00 \%  &	8.22 \% & -0.25 dB \\
\hline
\end{tabular}
\caption{Measure of speedup and \ac{bdrate} for \ac{abr} intra resolution analysis sharing with the state-of-the-art method}
\label{tab:intra_resolution_abr_result}
\end{table}

\section{Proposed Intra resolution analysis sharing in x265}
\label{sec:novel_intra_abr}

In this model, the \ac{cvbr} $B_{3}$ encode instance is the master representation. The subsequent encodes uses the analysis information from the master representation.

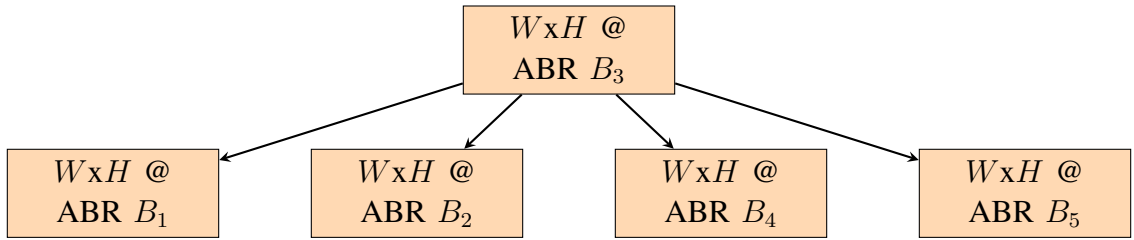
\begin{figure}[!htbp]
\begin{center}
\begin{tikzpicture}[node distance=1.0cm]
\node (i1) [process2] {$W$x$H$ @ ABR $B_{3}$};
\node (i2) [process2, below of=i1,xshift=-6cm, yshift = -0.9cm] {$W$x$H$ @ ABR $B_{1}$};
\node (i3) [process2, below of=i1,xshift=-2cm, yshift = -0.9cm] {$W$x$H$ @ ABR $B_{2}$};
\node (i4) [process2, below of=i1,xshift=2cm, yshift = -0.9cm] {$W$x$H$ @ ABR $B_{4}$};
\node (i5) [process2, below of=i1,xshift=6cm, yshift = -0.9cm] {$W$x$H$ @ ABR $B_{5}$};
\draw [arrow] (i1) -- (i2);
\draw [arrow] (i1) -- (i3);
\draw [arrow] (i1) -- (i4);
\draw [arrow] (i1) -- (i5);
\end{tikzpicture}
\end{center}
\caption{Intra resolution analysis sharing across $W$x$H$ resolution encode for \ac{abr} representation with the proposed method ($B_{1}$ > $B_{2} > $ $B_{3}$ > $B_{4}$ > $B_{5}$)}
\label{fig:intra_analysis_median_abr_diag}
\end{figure}

\noindent As shown in Table \ref{tab:intra_resolution_median_abr_result}, overall speedup for the five \ac{cvbr} representations for 540p videos is $56.14 \%$ with a \ac{bdrate} increase of $7.75 \%$ and \ac{bdpsnr} of $-0.27 dB$. For 1080p videos, the overall speedup is $59.15 \%$ with a \ac{bdrate} increase of $6.47 \%$ and \ac{bdpsnr} of $-0.18 dB$. For 2160p videos, the overall speedup is $60.09 \%$ with a \ac{bdrate} increase of $6.01 \%$ and \ac{bdpsnr} of $-0.17 dB$.

\begin{table}[!htbp]
\centering
\begin{tabular}{| p{2.0cm} | p{2.0cm}| p{2.0cm} |p{2.0cm} |}
\hline
\multicolumn{4}{|c|}{Test for \ac{abr} representations}\\
\hline
Resolution & $\Delta$T & \ac{bdrate} & \ac{bdpsnr}\\
\hline
\rowcolor{gray!10} 960x540 &  56.14 \%  &	7.75 \% & -0.27 dB \\
\rowcolor{blue!5} 1920x1080 &  59.15 \% & 6.47 \% & -0.18 dB \\
\rowcolor{gray!10} 3840x2160 &  60.09 \% &	6.01 \% & -0.17 dB \\
\hline
\end{tabular}
\caption{Measure of speedup and \ac{bdrate} for \ac{abr} intra resolution analysis sharing with the proposed method}
\label{tab:intra_resolution_median_abr_result}
\end{table}

\begin{figure}[!htbp]
    \centering
    \includegraphics[width=0.88\textwidth]{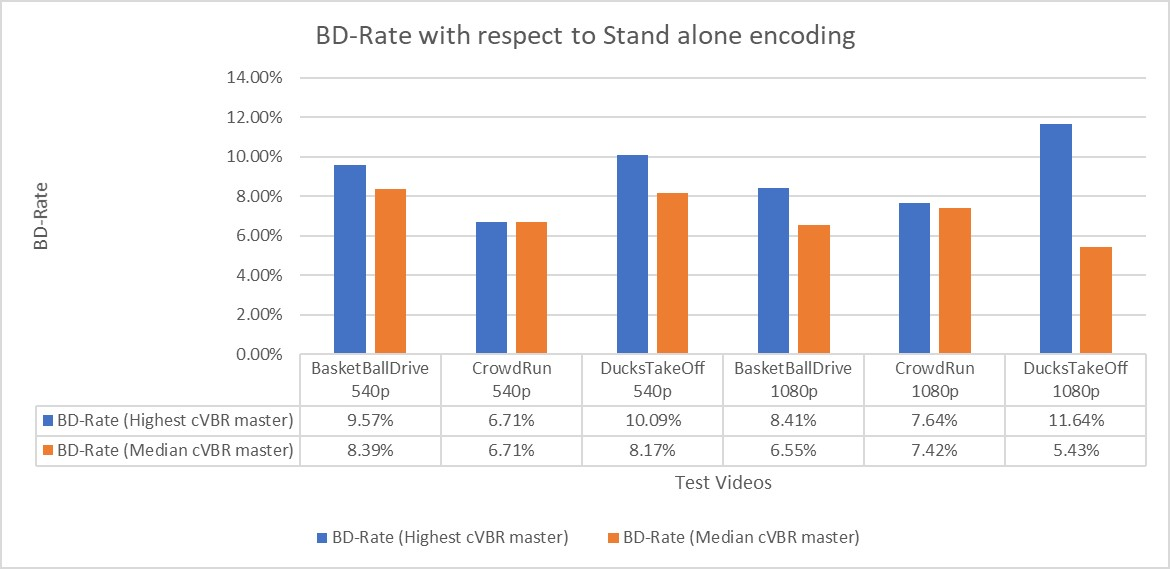}
    \caption{Comparison of \ac{bdrate} for \ac{cvbr} stand-alone encoding for the state-of-the-art method and the proposed method}
    \label{fig:bd_rate_abr_diag}
\end{figure}

\begin{figure}[!htbp]
    \centering
    \includegraphics[width=0.88\textwidth]{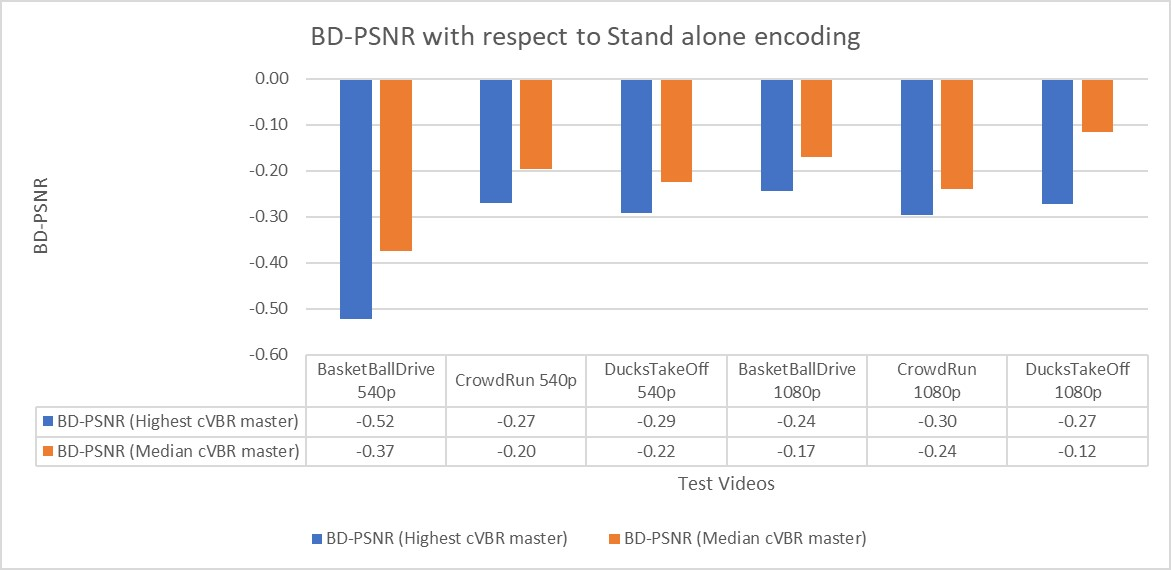}
    \caption{Comparison of \ac{bdpsnr} for \ac{cvbr} stand-alone encoding for the state-of-the-art method and the proposed method}
    \label{fig:bd_psnr_abr_diag}
\end{figure}

\section{Inter-resolution analysis sharing in x265}
\label{sec:inter_abr}
In this model, the \ac{cvbr} $B$ encode instance for $\frac{W}{2}$x$\frac{H}{2}$ resolution is the master representation. The subsequent encodes of $W$x$H$ resolution tier use the analysis information from the master representation. The results are evaluated for x265 analysis reuse level 10 mentioned in Table \ref{tab:x265_analysis}. Five \ac{cvbr} rates are used in $W$x$H$ resolution tier to calculate the \ac{bdrate} and \ac{bdpsnr} \cite{bd_rate_ref}.

\begin{figure}[!htbp]
\begin{center}
\begin{tikzpicture}[node distance=1.0cm]
\node (i1) [process4] {$\frac{W}{2}$x$\frac{H}{2}$ @ ABR $B$};
\node (i2) [process4, below of=i1,xshift=-5.4cm, yshift = -0.9cm] {$W$x$H$ @ ABR $B_{1}$};
\node (i3) [process4, below of=i1,xshift=-2.7cm, yshift = -1.3cm] {$W$x$H$ @ ABR $B_{2}$};
\node (i4) [process4, below of=i1,yshift = -1.3cm] {$W$x$H$ @ ABR $B_{3}$};
\node (i5) [process4, below of=i1,xshift=2.7cm, yshift = -1.3cm] {$W$x$H$ @ ABR $B_{4}$};
\node (i6) [process4, below of=i1,xshift=5.4cm, yshift = -0.9cm] {$W$x$H$ @ ABR $B_{5}$};
\draw [arrow] (i1) -- (i2);
\draw [arrow] (i1) -- (i3);
\draw [arrow] (i1) -- (i4);
\draw [arrow] (i1) -- (i5);
\draw [arrow] (i1) -- (i6);
\end{tikzpicture}
\end{center}
\caption{Inter resolution analysis sharing for \ac{abr} representation ($B_{1}$ > $B_{2} > $ $B_{3}$ > $B_{4}$ > $B_{5}$)}
\label{fig:inter_analysis_abr_diag}
\end{figure}
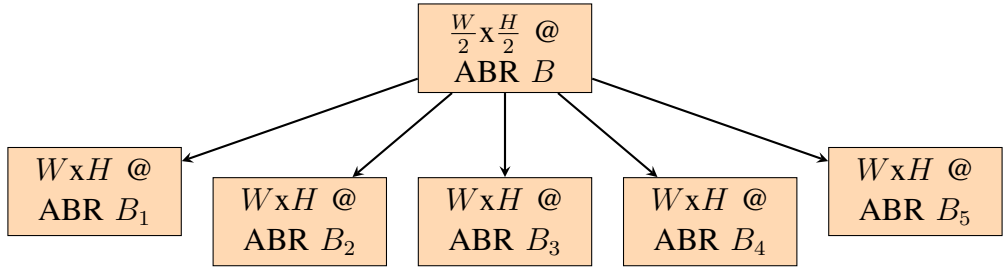

\noindent From Table \ref{tab:inter_resolution_2160p_median_abr_result} we see that, using 540p median \ac{cvbr} representation as the master encode, the 1080p resolution tier has a speedup of $65.64 \%$ with a \ac{bdrate} of $20.20 \%$ and \ac{bdpsnr} of $-0.64 dB$. Using 1080p median \ac{cvbr} representation as the master encode, the 2160p resolution tier has a speedup of $63.96 \%$ with a \ac{bdrate} of $25.91 \%$ and \ac{bdpsnr} of $-0.40 dB$. We notice that the \ac{bdrate} value is too high when we use all \ac{cu} analysis information without refinement. So, I use refinement options in x265 to have a reasonable \ac{bdrate}. The analysis refinement options in x265 were explained in Section \ref{sec:x265_analysis_refine_modes} where the intra-refinement options and inter-refinement options were explained in Table \ref{tab:intra_ref_x265} and Table \ref{tab:inter_ref_x265} respectively.\\

\begin{table}[!htbp]
\centering
\begin{tabular}{| p{2.0cm} | p{2.0cm}| p{2.0cm} |p{2.0cm} |}
\hline
\multicolumn{4}{|c|}{Test for \ac{abr} representations}\\
\hline
Resolution & $\Delta$T & \ac{bdrate} & \ac{bdpsnr}\\
\hline
\rowcolor{gray!10} 1920x1080 &  65.64 \%  & 20.20 \% & -0.64 dB \\
\rowcolor{blue!5} 3840x2160 &  63.96 \%  & 25.91 \% & -0.40 dB \\
\hline
\end{tabular}
\caption{Measure of speedup and \ac{bdrate} for 1080p and 2160p \ac{abr} inter-resolution analysis sharing with 540p median bitrate representation and 1080p median bitrate representation as master encode respectively}
\label{tab:inter_resolution_2160p_median_abr_result}
\end{table}

\begin{table}[!htbp]
\centering
\begin{tabular}{|p{2.0cm} | p{2.0cm}| p{2.0cm} |p{2.0cm} |}
\hline
\multicolumn{4}{|c|}{Test for \ac{abr} representations}\\
\hline
Resolution & $\Delta$T & \ac{bdrate} & \ac{bdpsnr}\\
\hline
\rowcolor{gray!10} 1920x1080 &  50.45 \%  & 8.62 \% & -0.21 dB \\
\rowcolor{blue!5} 3840x2160 &  52.82 \%  & 7.45 \% & -0.22 dB \\
\hline
\end{tabular}
\caption{Measure of speedup and \ac{bdrate} for 1080p and 2160p \ac{abr} inter-resolution analysis sharing with refinement with 540p median bitrate representation and 1080p median bitrate representation as master encode respectively}
\label{tab:inter_resolution_2160p_median_abr_refine_result}
\end{table}

\begin{figure}[!htbp]
    \centering
    \includegraphics[width=0.88\textwidth]{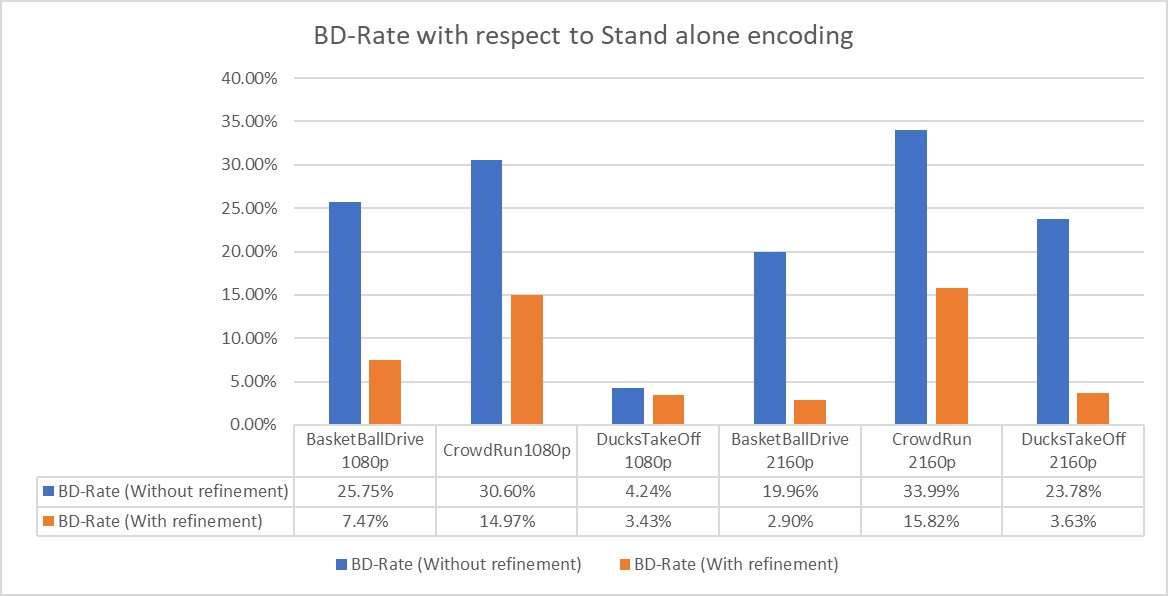}
    \caption{Comparison of \ac{bdrate} for \ac{cvbr} stand-alone encoding for inter-resolution with analysis refinement and without refinement}
    \label{fig:bd_rate_inter_res_refine}
\end{figure}

\begin{figure}[!htbp]
    \centering
    \includegraphics[width=0.88\textwidth]{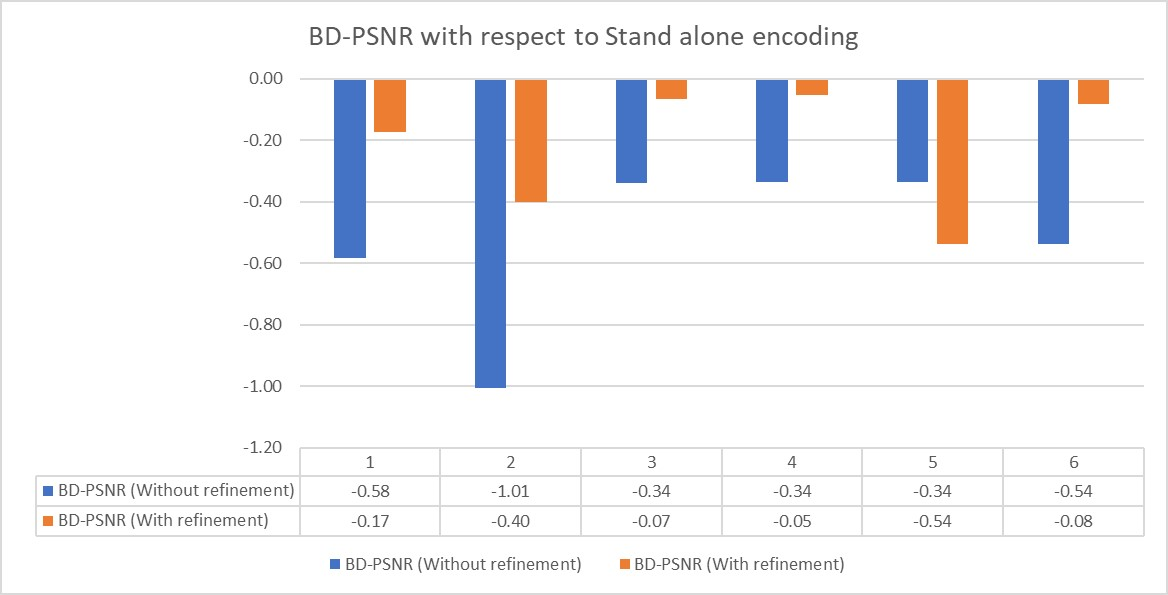}
    \caption{Comparison of \ac{bdpsnr} for \ac{cvbr} stand-alone encoding for inter-resolution with analysis refinement and without refinement}
    \label{fig:bd_psnr_inter_res_refine}
\end{figure}

\noindent From Table \ref{tab:inter_resolution_2160p_median_abr_refine_result} we see that, using 540p median \ac{cvbr} representation as the master encode, the 1080p resolution tier has a speedup of $50.45 \%$ with a \ac{bdrate} of $8.62 \%$ and \ac{bdpsnr} of $-0.21 dB$ with analysis refinement. Using 1080p median \ac{cvbr} representation as the master encode, the 2160p resolution tier has a speedup of $52.82 \%$ with a \ac{bdrate} of $7.45 \%$ and \ac{bdpsnr} of $-0.22 dB$ with analysis refinement. Figure \ref{fig:bd_rate_inter_res_refine} and Figure \ref{fig:bd_psnr_inter_res_refine} shows the graphical representation of the comparison of \ac{bdrate} and \ac{bdpsnr} for \ac{cvbr} stand-alone encoding for inter-resolution with analysis refinement and without refinement.

\newpage

\section{State of the art multi-encoding scheme}
\label{sec:stateofart_multi_res_rate}
Figure \ref{fig:stateofart_enc_analysis_share} shows the state-of-the-art multi-encoding scheme employed using the x265 HEVC encoder. The encoder analysis data from the 540p 3.5 Mbps \ac{cvbr} representation is used as the master representation to encode other 540p representations. The same data is refined to encode the 1080p 9.0 Mbps representation. In other words, the highest \ac{cvbr} representation of the lowest resolution tier is used to encode the remaining representations of the same tier and refined to encode the highest \ac{cvbr} representation of the subsequent resolution tier.   

\tikzstyle{process5} = [rectangle, minimum width=2.3cm, minimum height=0.75cm, text centered, text width=2.2cm, draw=black, fill=orange!30]
\begin{figure}[!htbp]
\begin{center}
\begin{tikzpicture}[node distance=1.0cm]
\node (i1) [process5] {540p @ 3.50Mbps};
\node (i2) [process5, below of=i1,xshift=-6.0cm, yshift = 0.1cm] {540p @ 3.00 Mbps};
\node (i3) [process5, below of=i1,xshift=-3.0cm, yshift = -0.6cm] {540p @ 2.50 Mbps};
\node (i4) [process5, below of=i1,yshift = -1.5cm] {1080p @ 9.00 Mbps};
\node (i5) [process5, below of=i1,xshift=3.0cm, yshift = -0.6cm] {540p @ 1.75 Mbps};
\node (i6) [process5, below of=i1,xshift=6.0cm, yshift = 0.1cm] {540p @ 1.00 Mbps};
\node (i7) [process5, below of=i4,xshift=-6.0cm, yshift = 0.1cm] {1080p @ 7.50 Mbps};
\node (i8) [process5, below of=i4,xshift=-3.0cm, yshift = -0.6cm] {1080p @ 5.50 Mbps};
\node (i9) [process5, below of=i4, yshift = -1.5cm] {2160p @ 19.00 Mbps};
\node (i10) [process5, below of=i4,xshift=3.0cm, yshift = -0.6cm] {1080p @ 4.50 Mbps};
\node (i11) [process5, below of=i4,xshift=6.0cm, yshift = 0.1cm] {1080p @ 3.50 Mbps};
\node (i12) [process5, below of=i9,xshift=-6.0cm, yshift = 0.1cm] {2160p @ 17.00 Mbps};
\node (i13) [process5, below of=i9,xshift=-3.0cm, yshift = -0.6cm] {2160p @ 15.00 Mbps};
\node (i14) [process5, below of=i9,xshift=3.0cm, yshift = -0.6cm] {2160p @ 13.00 Mbps};
\node (i15) [process5, below of=i9,xshift=6.0cm, yshift = 0.1cm] {2160p @ 11.00 Mbps};
\draw [arrow] (i1) -- (i2);
\draw [arrow] (i1) -- (i3);
\draw [arrow] (i1) -- (i4);
\draw [arrow] (i1) -- (i5);
\draw [arrow] (i1) -- (i6);
\draw [arrow] (i4) -- (i7);
\draw [arrow] (i4) -- (i8);
\draw [arrow] (i4) -- (i9);
\draw [arrow] (i4) -- (i10);
\draw [arrow] (i4) -- (i11);
\draw [arrow] (i9) -- (i12);
\draw [arrow] (i9) -- (i13);
\draw [arrow] (i9) -- (i14);
\draw [arrow] (i9) -- (i15);

\end{tikzpicture}
\end{center}
\caption{Representation of the state-of-the-art encoder analysis sharing across the resolutions and bitrates}
\label{fig:stateofart_enc_analysis_share}
\end{figure}
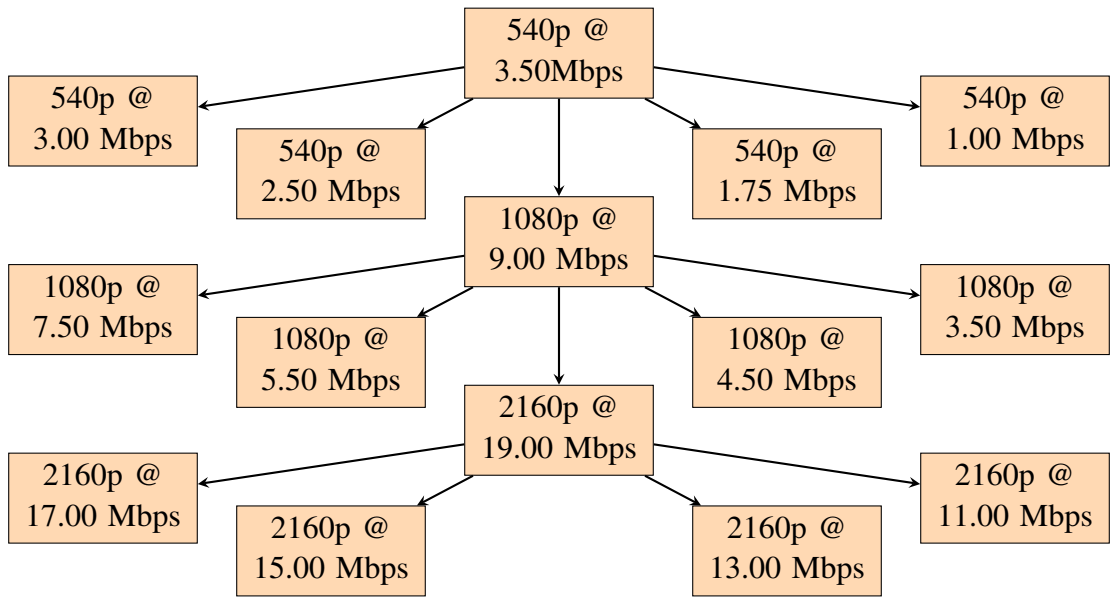

\section{Proposed multi-encoding scheme}
\label{sec:proposed_multi_res_rate}
\begin{figure}[!htbp]
\begin{center}
\begin{tikzpicture}[node distance=1.0cm]
\node (i1) [process5] {540p @ 2.50Mbps};
\node (i2) [process5, below of=i1,xshift=-6.0cm, yshift = 0.1cm] {540p @ 3.00 Mbps};
\node (i3) [process5, below of=i1,xshift=-3.0cm, yshift = -0.8cm] {540p @ 3.50 Mbps};
\node (i4) [process5, below of=i1,yshift = -1.5cm] {1080p @ 5.50 Mbps};
\node (i5) [process5, below of=i1,xshift=3.0cm, yshift = -0.8cm] {540p @ 1.75 Mbps};
\node (i6) [process5, below of=i1,xshift=6.0cm, yshift = 0.1cm] {540p @ 1.00 Mbps};
\node (i7) [process5, below of=i4,xshift=-6.0cm, yshift = 0.1cm] {1080p @ 7.50 Mbps};
\node (i8) [process5, below of=i4,xshift=-3.0cm, yshift = -0.8cm] {1080p @ 9.00 Mbps};
\node (i9) [process5, below of=i4, yshift = -1.5cm] {2160p @ 15.00 Mbps};
\node (i10) [process5, below of=i4,xshift=3.0cm, yshift = -0.8cm] {1080p@ 4.50 Mbps};
\node (i11) [process5, below of=i4,xshift=6.0cm, yshift = 0.1cm] {1080p@ 3.50 Mbps};
\node (i12) [process5, below of=i9,xshift=-6.0cm, yshift = 0.1cm] {2160p@ 17.00 Mbps};
\node (i13) [process5, below of=i9,xshift=-3.0cm, yshift = -0.8cm] {2160p@ 19.00 Mbps};
\node (i14) [process5, below of=i9,xshift=3.0cm, yshift = -0.8cm] {2160p@ 13.00 Mbps};
\node (i15) [process5, below of=i9,xshift=6.0cm, yshift = 0.1cm] {2160p@ 11.00 Mbps};
\draw [arrow] (i1) -- (i2);
\draw [arrow] (i1) -- (i3);
\draw [arrow] (i1) -- (i4);
\draw [arrow] (i1) -- (i5);
\draw [arrow] (i1) -- (i6);
\draw [arrow] (i4) -- (i7);
\draw [arrow] (i4) -- (i8);
\draw [arrow] (i4) -- (i9);
\draw [arrow] (i4) -- (i10);
\draw [arrow] (i4) -- (i11);
\draw [arrow] (i9) -- (i12);
\draw [arrow] (i9) -- (i13);
\draw [arrow] (i9) -- (i14);
\draw [arrow] (i9) -- (i15);
\end{tikzpicture}
\end{center}
\caption{Representation of the proposed encoder analysis sharing across the resolutions and bitrates}
\label{fig:proposed_enc_analysis_share}
\end{figure}
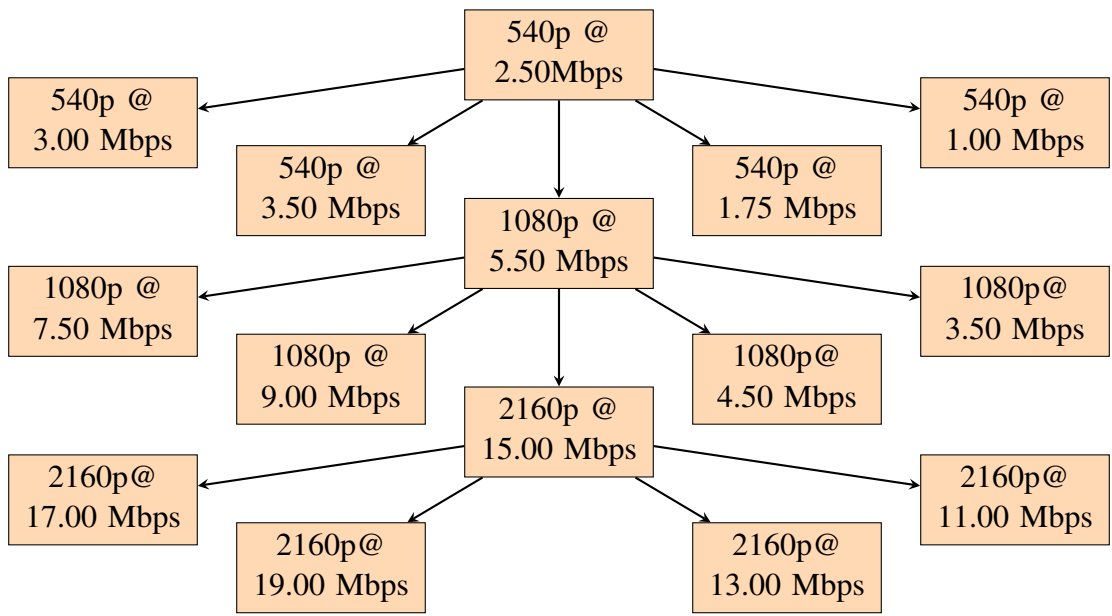

Since an optimal intra-resolution and inter-resolution analysis sharing scheme is found, a novel multi-encoding scheme is proposed to generate \ac{cvbr} bit-streams for various resolutions and bitrates for \ac{abr} streaming. In Figure \ref{fig:proposed_enc_analysis_share}, the proposed multi-encoding method is depicted with the optimal analysis-sharing scheme. The 540p 2.5 Mbps \ac{cvbr} representation is the master encode representation. The encoder analysis information is shared with the other 540p resolution representations. It is also shared with the 1080p 5.5 Mbps representation, which refines the analysis data and reuses it. The analysis information from 1080p 5.5 Mbps representation encodes other 1080p resolution representations and 2160p 15 Mbps \ac{cvbr} representation. The proposed method gives an average 7\% less \ac{bdrate} and 9\% more speedup compared to the state-of-the-art scheme.

\section{Improvement in XVC encoding with SIMD optimization}
\label{sec:xvc_simd_res}
The experiments are focused on all presets of the Arowana XVC encoder to represent the wide set of use cases that XVC can be used for since these presets represent a wide variety of trade-offs between compression efficiency and encoding speed (measured in frames per second). The 'slow' preset generates the most efficient encode but is the slowest; this preset is also the preferred choice for offline encoding use cases. The 'superultrafast' preset is the quickest setting of Arowana XVC but generates the encode with the lowest efficiency. The 'veryfast' preset represents an intermediate trade-off between performance and encoder efficiency. Typically, the more efficient presets employ more tools of XVC, resulting in more compute per pixel than the less efficient presets.\\

\begin{figure}[!ht]
    \centering
    \includegraphics[width=0.95\textwidth]{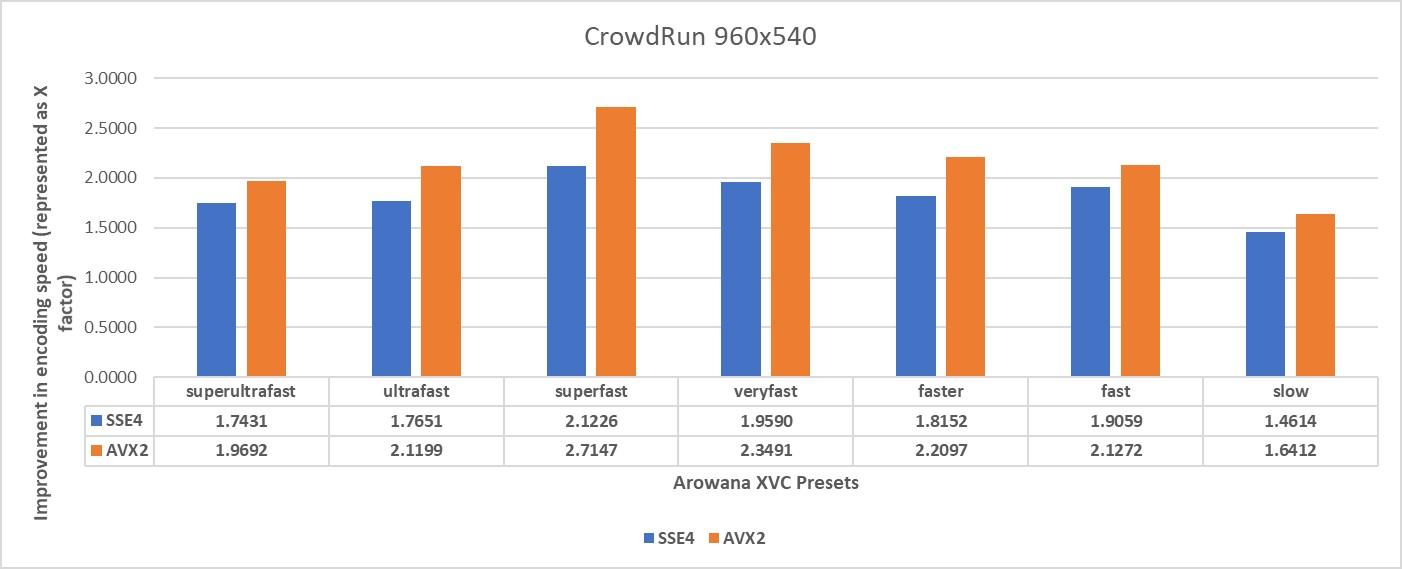}
    \caption{XVC encoding speedup for CrowdRun video for Arowana XVC presets (GOP length=2 seconds)}
    \label{fig:speedup1}
\end{figure}

\begin{figure}[!ht]
    \centering
    \includegraphics[width=0.95\textwidth]{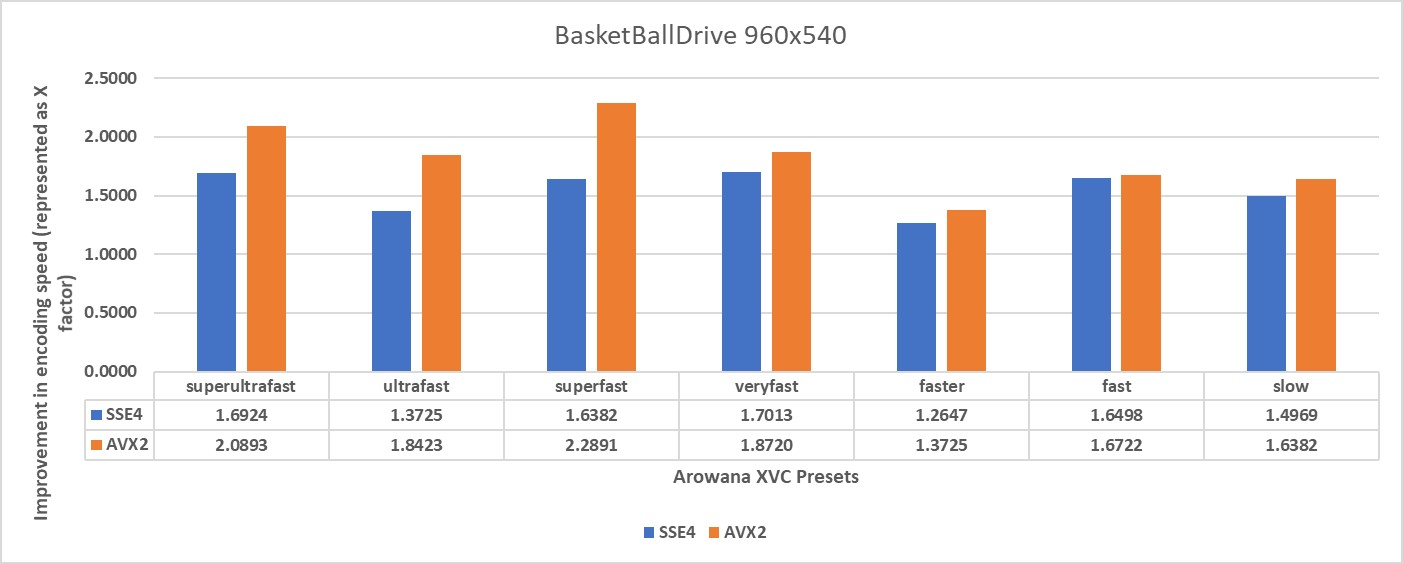}
    \caption{XVC encoding speedup for BasketBallDrive video  for Arowana XVC presets (GOP length=2 seconds)}
    \label{fig:speedup2}
\end{figure}

\begin{figure}[!ht]
    \centering
    \includegraphics[width=0.95\textwidth]{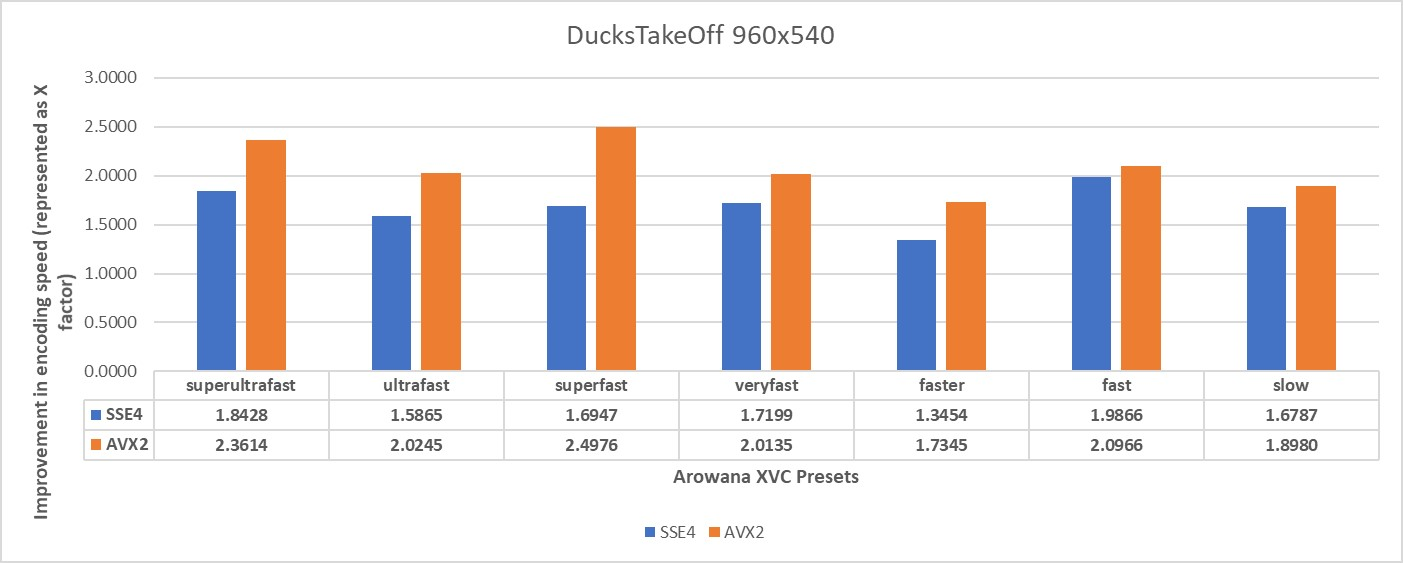}
    \caption{XVC encoding speedup for DucksTakeOff video  for Arowana XVC presets (GOP length=2 seconds)}
    \label{fig:speedup3}
\end{figure}

\noindent Figure \ref{fig:speedup1}, Figure \ref{fig:speedup2} and Figure \ref{fig:speedup3} shows the improvement in the encoding speed of Arowana XVC encoder for the given test videos at 540p resolution for all the presents defined in Arowana encoder. We notice that we have a net speedup of about 2.5x in some cases. This demonstrates that optimizing frequently used functions in the video encoder can boost the encoding speed significantly. The results are mainly obtained from optimizing the cost and interpolation filter functions. The critical functions like transforms are not yet optimized. The more the functions are vectorized, the speedup also increases.
\newpage

%% file: kth_chapter7_conclusion.tex
\chapter{Conclusions and Future Directions}
\label{sec:conc_future}
This chapter concludes the project report and summarises the key findings during the project. Section \ref{sec:conclusion} discusses the conclusions and Section \ref{sec:future} discusses the future directions of this work which haven't been explored as part of this project respectively.

\section{Conclusions}
\label{sec:conclusion}
The state-of-the-art intra-resolution muti-rate \ac{hevc} encoding scheme is modeled and evaluated in this project. The highest bitrate representation is used as the master representation. To mitigate the performance bottleneck in parallel encoding and improve the overall compression efficiency, a novel scheme is proposed for intra-resolution muti-rate \ac{hevc} encoding, where the median bitrate representation is used as the master representation. The state-of-the-art multi-resolution \ac{hevc} encoding scheme with inter-resolution analysis sharing is evaluated. With the updated optimized schemes, a multi-encoding scheme with multi-resolution tier multi-rate \ac{hevc} encoding scheme for \ac{abr} streaming is proposed. In the proposed scheme, the encoding instances of the same resolution tier reuse the encoder decisions of the encoding instance with median \ac{cvbr}. The encoder decisions are refined and reused in the subsequent resolution tier encoding instance with median \ac{cvbr}. The evaluations of the schemes were conducted on the x265 \ac{hevc} encoder. Experimental results demonstrate significant speedups with a marginal impact on coding efficiency.\\

\noindent In this project, the impact of \ac{simd} extension optimizations in the encoding speeds of the next-generation video codecs is evaluated. The optimizations are implemented in the Arowana XVC encoder, where the encoding speeds were accelerated  up to 2.7x by vectorizing some essential cost functions and interpolation filter functions. It is also noticed that the handwritten assembly is the best method to get the best performance boost, though it needs a lot of human effort.

\section{Future Directions}
\label{sec:future}
In this project, the results for the parallel encoding of the various encoding instances are not considered and discussed due to hardware constraints. In the proposed model for multi-encoding discussed in Section \ref{sec:proposed_multi_res_rate}, other representations of $\frac{W}{4}$x$\frac{H}{4}$ resolution tier can be run with a frame delay parallel to the $\frac{W}{4}$x$\frac{H}{4}$ median \ac{cvbr} representation. Also, the median \ac{cvbr} representation of $\frac{W}{2}$x$\frac{H}{2}$ can also be run with a frame delay parallel to the $\frac{W}{4}$x$\frac{H}{4}$ median \ac{cvbr} representation. This encoding structure is possible for high-end servers with multiple \ac{cpu} cores and threads. The parallel encoding approach can scale the performance benefits mentioned in the project.\\

\noindent In this project, only a few critical functions in the Arowana XVC encoder are optimized. There are still many vectorizable codes in Arowana, like DCT, iDCT, transpose, and other functions that can significantly boost the encoder speed across all presets. According to \cite{simd_ref6}, x265 received a 5.5x speedup after implementing optimization of all generations of Intel \ac{simd} till \ac{avx2}. Arowana being a more complex encoder can achieve a better performance boost with full-scale optimizations. The performance improvement in high-end servers can be further scaled up by \ac{avx512} optimizations.

%% file: kth_appendix1_x265_options.tex
\chapter{x265 CLI Options}
\label{sec:x265_cli}

\section{Preview of x265 encoder analysis share modes}
\label{sec:x265_analysis_modes}
The x265 command line parameter \textit{--analysis-save} is used to generate the corresponding encoder analysis information as a metadata file from the master/ base encodes. In the master/ base encode, an additional parameter \textit{--analysis-save-reuse-level} defines the extent of the saved information in the metadata file as shown in Table \ref{tab:x265_analysis}. Subsequently, the dependent representation reuses this information in the metadata file by combining the \textit{--analysis-load} and \textit{--analysis-load-reuse-level} parameters. The dependent representation can either be of the exact resolution as that of the master/ base, albeit targeting a different quality level, or have a different resolution from the master/ base. \textit{--scale-factor} option is to scale up the encoder analysis data to the required resolution. Currently, x265 allows this kind of analysis sharing only across dyadic resolutions (i.e., multiples of both the width and height by a factor of two). The shared analysis data includes the encoder decisions like quad-tree structure, prediction modes (e.g., intra/inter prediction modes), motion vectors (MVs), and frame-level information, such as slice-types, reference lists, and distortion information. The choice of the analysis share level depends on the trade-off between compression efficiency and speedup. In this work, I consider only levels- 4, 6, and 10.

\begin{table}[!htbp]
\centering
\begin{tabular}{ |p{3.0cm}|p{10.8cm}|}
\hline
\multicolumn{2}{|c|}{x265 Analysis reuse levels}\\
\hline
Level & Description\\
\hline
\rowcolor{blue!5} 1 & Reuse Lookahead information \\
\rowcolor{gray!10} 2-4 & Level 1 + Reuse intra/inter modes, ref’s \\
\rowcolor{blue!5} 5-6 & Level 2 + Reuse rect-amp \\
\rowcolor{gray!10} 7 & Level 5 + Reuse AVC size CU refinement \\
\rowcolor{blue!5} 8-9 & Level 5 + Reuse AVC size Full CU analysis-info \\
\rowcolor{gray!10} 10 & Level 5 + Reuse Full CU analysis-info \\
\hline
\end{tabular}
\caption{x265 analysis reuse levels and their respective descriptions \cite{x265_cli_ref}}
\label{tab:x265_analysis}
\end{table}

\section{Preview of x265 encoder analysis refinement modes}
\label{sec:x265_analysis_refine_modes}
For multi-resolution encoding, there are specific refinement techniques used in x265. The CLI option \textit{--refine-intra} is used to enable refinement of intra blocks in the current representation. The various values for the parameter and the respective functionality are shown in Table \ref{tab:intra_ref_x265}. 

\begin{table}[!htbp]
\centering
\begin{tabular}{ |p{3.0cm}|p{10.8cm}|}
\hline
\multicolumn{2}{|c|}{Intra Refinement}\\
\hline
Level & Description \\
\hline
\rowcolor{gray!10} 0 & Forces both mode and depth from the reference encode \\
\rowcolor{blue!5} 1 & Evaluates all intra modes at current depth(n) and at depth (n+1) when current block size is one greater than the min-cu-size and forces modes for larger blocks. \\
\rowcolor{gray!10} 2 & In addition to the functionality of level 1, at all depths, force \newline
	(i) only depth when the angular mode is chosen by the save encode.\newline
	(ii) depth and mode when other intra-modes are chosen by the save encode. \\
\rowcolor{blue!5} 3 & Perform analysis of intra modes for depth reused from reference encode \\
\rowcolor{gray!10} 4 & Does not reuse any analysis information - redo analysis for the intra block \\
\hline
\end{tabular}
\caption{Intra refinement techniques in x265 \cite{x265_cli_ref}}
\label{tab:intra_ref_x265}
\end{table}

\noindent The CLI option \textit{--refine-inter} is used to enable refinement of inter blocks in the current representation. The various values for the parameter and the respective functionality are shown in Table \ref{tab:inter_ref_x265}. The CLI option \textit{--refine-mv} is used to enable the refinement of the motion vector for the scaled video. The best motion vector is evaluated based on the level selected. The various levels and the respective functionality is shown in Table \ref{tab:mv_refine_x265}.\\

\begin{table}[!htbp]
\centering
\begin{tabular}{ |p{3.0cm}|p{10.8cm}|}
\hline
\multicolumn{2}{|c|}{Inter Refinement}\\
\hline
Level & Description \\
\hline
\rowcolor{gray!10} 0 & Forces both mode and depth from the reference encode \\
\rowcolor{blue!5} 1 & Evaluates all inter modes at current depth(n) and at depth (n+1) when current block size is one greater than the min-cu-size and forces modes for larger blocks. \\
\rowcolor{gray!10} 2 & In addition to the functionality of level 1, restricts the modes 
	evaluated when specific modes are decided as the best mode by the save encode \newline
		2Nx2N in save encode - disable re-evaluation of rect and amp \newline
    	SKIP in save encode  - re-evaluates only SKIP, MERGE, and 2Nx2N modes.
	\\
\rowcolor{blue!5} 3 & Perform analysis of inter modes while reusing depths from reference encode \\
\hline
\end{tabular}
\caption{Inter refinement techniques in x265 \cite{x265_cli_ref}}
\label{tab:inter_ref_x265}
\end{table}

\begin{table}[!htbp]
\centering
\begin{tabular}{ |p{3.0cm}|p{10.8cm}|}
\hline
\multicolumn{2}{|c|}{Motion Vector Refinement}\\
\hline
Level & Description \\
\hline
\rowcolor{gray!10} 1 & Search around scaled MV \\
\rowcolor{blue!5} 2 & Level 1 + Search around best AMVP candidate \\
\rowcolor{gray!10} 3 & Level 2 + Search around the other AMVP candidate \\
\hline
\end{tabular}
\caption{Motion vector refinement techniques in x265 \cite{x265_cli_ref}}
\label{tab:mv_refine_x265}
\end{table}

%% file: kth_appendix2_sse4.tex
\chapter{SSE4 SIMD optimization}
\label{sec:sse4_test_res}
\begin{table}
\fontsize{8pt}{10pt}\selectfont
\begin{tabular}{| p{3.3cm}| p{0.7cm}|  p{3.3cm}|  p{0.7cm}|  p{3.3cm}|  p{0.7cm}| }
\hline
Primitive & IPC \newline Gain & Primitive & IPC \newline Gain & Primitive & IPC \newline Gain\\
\hline
                   sad\_ac\_p[1][8x4] & 0.84 &                   sad\_ac\_p[1][64x8] & 1.2 &                sad\_ac\_res[1][64x16] & 1.36\\
                mean\_diff\_p[1][4x8] & 0.9 &                 subtract\_res[32x16] & 1.2 &                    sub\_weight[32x8] & 1.38\\
                subtract\_pix[64x64] & 0.92 &                 sad\_ac\_res[0][64x8] & 1.21 &                 sad\_ac\_res[0][16x8] & 1.4\\
                subtract\_pix[32x32] & 0.95 &                   sad\_ac\_p[0][16x2] & 1.22 &                sad\_ac\_res[1][16x32] & 1.4\\
                subtract\_res[64x32] & 0.95 &                 sad\_ac\_res[0][32x4] & 1.22 &                  sad\_ac\_res[0][8x4] & 1.41\\
                   sad\_ac\_p[1][8x2] & 0.98 &                 subtract\_pix[64x16] & 1.23 &                 sad\_ac\_res[1][32x8] & 1.41\\
                 copy\_bipred[64x32] & 1.02 &                   sub\_weight[64x32] & 1.23 &                 mean\_diff\_p[1][8x2] & 1.42\\
\hline
\end{tabular}
\end{table}
\newpage
\begin{table}
\fontsize{8pt}{10pt}\selectfont
\begin{tabular}{| p{3.3cm}| p{0.7cm}|  p{3.3cm}|  p{0.7cm}|  p{3.3cm}|  p{0.7cm}| }
\hline
Primitive & IPC \newline Gain & Primitive & IPC \newline Gain & Primitive & IPC \newline Gain\\
\hline
                 copy\_bipred[32x64] & 1.1 &                sad\_ac\_res[0][32x64] & 1.28 &                 sad\_ac\_res[0][8x64] & 1.46\\
                 subtract\_pix[32x8] & 1.1 &                  subtract\_pix[64x8] & 1.28 &                sad\_ac\_res[1][64x64] & 1.46\\
                 sad\_ac\_p[1][16x32] & 1.11 &                 sad\_ac\_res[0][16x2] & 1.29 &                 sad\_ac\_res[1][8x32] & 1.48\\
              mean\_diff\_res[1][4x2] & 1.12 &                  sad\_ac\_res[0][8x2] & 1.3 &                 sad\_ac\_res[1][32x4] & 1.48\\
                 sad\_ac\_res[1][8x2] & 1.12 &                sad\_ac\_res[0][32x32] & 1.3 &                sad\_ac\_res[1][32x16] & 1.48\\
                subtract\_res[64x64] & 1.12 &                    sad\_ac\_p[1][8x8] & 1.31 &                  sad\_ac\_p[1][32x32] & 1.48\\
                  sad\_ac\_p[1][16x2] & 1.14 &                   sub\_weight[32x16] & 1.31 &                sad\_ac\_res[1][32x32] & 1.48\\
\hline
\end{tabular}
\end{table}
\newpage
\begin{table}
\fontsize{8pt}{10pt}\selectfont
\begin{tabular}{| p{3.3cm}| p{0.7cm}|  p{3.3cm}|  p{0.7cm}|  p{3.3cm}|  p{0.7cm}| }
\hline
Primitive & IPC \newline Gain & Primitive & IPC \newline Gain & Primitive & IPC \newline Gain\\
\hline

                 sad\_ac\_res[0][4x8] & 2.09 &                    calc\_sign[16x16] & 2.28 &                        sad\_pp[4x16] & 2.5\\
                   sub\_weight[16x8] & 2.09 &                        sad\_pp[16x2] & 2.32 &                         sad\_pp[8x2] & 2.5\\
                      sad\_sp[32x64] & 2.11 &                        sad\_pp[32x8] & 2.33 &                mean\_diff\_p[0][32x4] & 2.5\\
                  subtract\_res[4x8] & 2.12 &                   sub\_weight[16x16] & 2.33 &                 subtract\_pix[16x32] & 2.51\\
                  sub\_weight[16x64] & 2.13 &                      sad\_sp\_x3[8x8] & 2.35 &                        sad\_sp[4x32] & 2.52\\
                 subtract\_res[16x4] & 2.15 &                  subtract\_pix[16x8] & 2.37 &                    sad\_sp\_x3[16x32] & 2.52\\
                subtract\_res[16x16] & 2.15 &                         sad\_sp[8x4] & 2.39 &                       sad\_sp[64x32] & 2.52\\
\hline
\end{tabular}
\end{table}
\newpage
\begin{table}
\fontsize{8pt}{10pt}\selectfont
\begin{tabular}{| p{3.3cm}| p{0.7cm}|  p{3.3cm}|  p{0.7cm}|  p{3.3cm}|  p{0.7cm}| }
\hline
Primitive & IPC \newline Gain & Primitive & IPC \newline Gain & Primitive & IPC \newline Gain\\
\hline
                 chroma\_vps\_10[4x2] & 2.99 &             mean\_diff\_res[1][32x64] & 3.16 &             mean\_diff\_res[1][64x32] & 3.32\\
                        sad\_pp[8x8] & 3 &                    sub\_weight[4x16] & 3.17 &                  chroma\_vps\_12[8x4] & 3.32\\
            mean\_diff\_res[1][16x64] & 3 &             mean\_diff\_res[0][32x64] & 3.18 &             luma\_inv\_quant\_10b[4x4] & 3.32\\
                 chroma\_vss\_10[8x8] & 3 &                  chroma\_vps\_10[8x2] & 3.18 &                   rdo\_clipmul[4x16] & 3.32\\
                    sub\_weight[4x8] & 3 &               rdo\_sign\_reapply[4x8] & 3.18 &              rdo\_sign\_reapply[4x16] & 3.32\\
             mean\_diff\_res[0][16x8] & 3.01 &                   rdo\_clipmul[8x32] & 3.18 &                  chroma\_vsp\_12[8x2] & 3.33\\
               mean\_diff\_p[1][32x8] & 3.01 &               mean\_diff\_p[0][32x32] & 3.19 &                      calc\_sign[8x2] & 3.33\\
\hline
\end{tabular}
\end{table}
\newpage
\begin{table}
\fontsize{8pt}{10pt}\selectfont
\begin{tabular}{| p{3.3cm}| p{0.7cm}|  p{3.3cm}|  p{0.7cm}|  p{3.3cm}|  p{0.7cm}| }
\hline
Primitive & IPC \newline Gain & Primitive & IPC \newline Gain & Primitive & IPC \newline Gain\\
\hline
                 chroma\_vpp\_12[8x4] & 4.2 &           luma\_inv\_quant\_10b[16x64] & 4.56 &             rdo\_sign\_reapply[16x64] & 4.77\\
                 chroma\_vpp\_12[4x8] & 4.23 &                  chroma\_vsp\_10[8x2] & 4.57 &                    add\_avg\_10[8x16] & 4.79\\
                   rdo\_clipmul[8x4] & 4.24 &                 chroma\_vsp\_10[8x16] & 4.57 &                 chroma\_vss\_10[8x64] & 4.79\\
                 chroma\_vss\_12[4x8] & 4.25 &                ang\_fast\_copy[16x64] & 4.57 &           luma\_inv\_quant\_12b[16x64] & 4.8\\
                  add\_clip\_10[4x32] & 4.25 &            luma\_inv\_quant\_10b[8x32] & 4.57 &                 chroma\_vpp\_12[4x32] & 4.82\\
                    add\_avg\_12[8x2] & 4.26 &                        ssd\_sp[4x16] & 4.58 &                       satd\_sp[8x16] & 4.84\\
                  add\_clip\_12[4x32] & 4.27 &                        satd\_sp[8x8] & 4.58 &                       satd\_pp[8x32] & 4.84\\
\hline
\end{tabular}
\end{table}

\newpage
\begin{table}
\fontsize{8pt}{10pt}\selectfont
\begin{tabular}{| p{3.3cm}| p{0.7cm}|  p{3.3cm}|  p{0.7cm}|  p{3.3cm}|  p{0.7cm}| }
\hline
Primitive & IPC \newline Gain & Primitive & IPC \newline Gain & Primitive & IPC \newline Gain\\
\hline
                     satd\_sp[64x16] & 5.63 &                ang\_fast\_copy[32x64] & 5.81 &                 chroma\_vsp\_12[16x8] & 5.99\\
               chroma\_vps\_12[64x16] & 5.63 &                       satd\_sp[16x8] & 5.83 &                chroma\_vpp\_12[64x64] & 5.99\\
                ang\_fast\_copy[32x8] & 5.64 &           luma\_inv\_quant\_12b[32x32] & 5.83 &                    luma\_vsp\_10[8x4] & 5.99\\
                        ssd\_pp[8x4] & 5.65 &                         ssd\_ss[4x8] & 5.84 &                     calc\_sign[8x16] & 5.99\\
                chroma\_vpp\_12[4x16] & 5.65 &                    luma\_vps\_12[8x8] & 5.84 &                  luma\_vsp\_12[16x32] & 6\\
          luma\_inv\_quant\_12b[32x16] & 5.65 &                   luma\_vsp\_12[8x32] & 5.84 &            luma\_inv\_quant\_12b[32x4] & 6\\
          luma\_inv\_quant\_12b[64x32] & 5.66 &              rdo\_sign\_reapply[32x8] & 5.84 &                     calc\_sign[8x64] & 6\\
\hline
\end{tabular}
\end{table}
\newpage
\begin{table}
\fontsize{8pt}{10pt}\selectfont
\begin{tabular}{| p{3.3cm}| p{0.7cm}|  p{3.3cm}|  p{0.7cm}|  p{3.3cm}|  p{0.7cm}| }
\hline
Primitive & IPC \newline Gain & Primitive & IPC \newline Gain & Primitive & IPC \newline Gain\\
\hline
                 add\_clip\_12[16x32] & 6.85 &                   luma\_vps\_12[4x32] & 7.03 &                        ssd\_pp[32x4] & 7.27\\
                   add\_avg\_12[16x8] & 6.86 &                  luma\_vsp\_12[64x32] & 7.03 &                   luma\_vss\_10[8x16] & 7.27\\
                chroma\_vsp\_10[64x8] & 6.86 &                chroma\_vpp\_10[16x32] & 7.05 &                 ang\_fast\_copy[16x8] & 7.27\\
                ang\_fast\_copy[64x8] & 6.86 &                   luma\_vps\_10[4x16] & 7.05 &                 chroma\_vps\_10[64x8] & 7.29\\
                     satd\_pp[16x32] & 6.87 &                 chroma\_vpp\_10[8x64] & 7.07 &                   copy\_bipred[8x32] & 7.32\\
               chroma\_vpp\_10[32x16] & 6.87 &                   luma\_vsp\_10[16x8] & 7.07 &                   luma\_vpp\_12[4x32] & 7.32\\
                chroma\_vpp\_10[32x8] & 6.88 &                chroma\_vsp\_10[64x16] & 7.08 &                  add\_clip\_10[16x16] & 7.33\\
\hline
\end{tabular}
\end{table}
\newpage
\begin{table}
\fontsize{8pt}{10pt}\selectfont
\begin{tabular}{| p{3.3cm}| p{0.7cm}|  p{3.3cm}|  p{0.7cm}|  p{3.3cm}|  p{0.7cm}| }
\hline
Primitive & IPC \newline Gain & Primitive & IPC \newline Gain & Primitive & IPC \newline Gain\\
\hline
               chroma\_vss\_12[64x32] & 8.49 &                  luma\_vss\_12[32x64] & 8.74 &              rdo\_num\_nonzero[32x32] & 9.62\\
               chroma\_vps\_12[64x64] & 8.49 &                  add\_clip\_12[64x16] & 8.77 &                  ang\_fast\_copy[8x8] & 9.65\\
                 luma\_vps\_10[32x64] & 8.49 &                  luma\_vss\_12[32x16] & 8.83 &                   add\_clip\_12[64x8] & 9.7\\
                  block\_zero[16x32] & 8.5 &                   add\_avg\_12[32x16] & 8.89 &                  luma\_vss\_10[32x32] & 9.71\\
                  luma\_vpp\_10[32x8] & 8.51 &                   luma\_vpp\_10[64x8] & 8.93 &                   add\_clip\_10[64x8] & 9.75\\
               chroma\_vpp\_10[64x64] & 8.52 &                  luma\_vpp\_12[32x32] & 8.98 &              rdo\_num\_nonzero[64x32] & 9.77\\
                       ssd\_pp[32x8] & 8.53 &                   luma\_vss\_12[64x8] & 9.02 &                     block\_zero[8x2] & 9.86\\
\hline
\end{tabular}
\end{table}

%% file: kth_appendix3_avx2.tex
\chapter{AVX2 SIMD optimization}
\label{sec:avx2_test_res}
\begin{table}
\fontsize{8pt}{10pt}\selectfont
\begin{tabular}{| p{3.3cm}| p{0.7cm}|  p{3.3cm}|  p{0.7cm}|  p{3.3cm}|  p{0.7cm}| }
\hline
Primitive & IPC \newline Gain & Primitive & IPC \newline Gain & Primitive & IPC \newline Gain\\
\hline
                            dst4[2] & 1.02 &                    sad\_ac\_p[0][8x4] & 1.75 &                sad\_ac\_res[0][64x32] & 1.96\\
                            dst4[1] & 1.04 &                   sad\_ac\_p[0][16x4] & 1.77 &                 subtract\_res[32x64] & 1.97\\
                   sad\_ac\_p[1][8x2] & 1.22 &                sad\_ac\_res[1][64x16] & 1.77 &                sad\_ac\_res[1][32x32] & 1.99\\
              mean\_diff\_res[1][8x2] & 1.24 &                   sad\_ac\_p[0][32x4] & 1.79 &                 sad\_ac\_res[1][8x16] & 2\\
                sad\_ac\_res[1][16x2] & 1.26 &                   copy\_bipred[32x4] & 1.79 &                sad\_ac\_res[1][64x64] & 2\\
                 sad\_ac\_res[0][8x4] & 1.35 &                 subtract\_res[32x16] & 1.79 &                  subtract\_pix[32x8] & 2\\
                   sad\_ac\_p[0][8x2] & 1.4 &                 sad\_ac\_res[1][64x8] & 1.81 &                   sub\_weight[32x16] & 2\\
\hline
\end{tabular}
\end{table}
\newpage
\begin{table}
\fontsize{8pt}{10pt}\selectfont
\begin{tabular}{| p{3.3cm}| p{0.7cm}|  p{3.3cm}|  p{0.7cm}|  p{3.3cm}|  p{0.7cm}| }
\hline
Primitive & IPC \newline Gain & Primitive & IPC \newline Gain & Primitive & IPC \newline Gain\\
\hline
                  sad\_ac\_p[0][16x2] & 1.62 &                  sad\_ac\_res[0][8x2] & 1.91 &                  sad\_ac\_p[1][32x64] & 2.12\\
                  sad\_ac\_p[1][16x4] & 1.62 &                  sad\_ac\_res[1][8x4] & 1.91 &                  subtract\_res[32x8] & 2.12\\
                 sad\_ac\_p[0][32x64] & 1.62 &                   sad\_ac\_p[0][16x8] & 1.91 &                sad\_ac\_res[1][32x64] & 2.13\\
                 copy\_bipred[64x64] & 1.62 &                   sad\_ac\_p[1][64x8] & 1.91 &                   sub\_weight[64x32] & 2.13\\
                 subtract\_pix[32x4] & 1.62 &                sad\_ac\_res[1][64x32] & 1.91 &                 sad\_ac\_res[1][32x8] & 2.14\\
                 copy\_bipred[32x16] & 1.63 &                 subtract\_pix[32x64] & 1.91 &                   sub\_weight[64x16] & 2.14\\
                 sad\_ac\_p[1][64x64] & 1.65 &                sad\_ac\_res[1][16x16] & 1.92 &                 sad\_ac\_res[0][32x8] & 2.15\\
\hline
\end{tabular}
\end{table}
\newpage
\begin{table}
\fontsize{8pt}{10pt}\selectfont
\begin{tabular}{| p{3.3cm}| p{0.7cm}|  p{3.3cm}|  p{0.7cm}|  p{3.3cm}|  p{0.7cm}| }
\hline
Primitive & IPC \newline Gain & Primitive & IPC \newline Gain & Primitive & IPC \newline Gain\\
\hline
                   sad\_sp\_x3[32x64] & 3.48 &             luma\_inv\_quant\_10b[8x4] & 3.84 &                mean\_diff\_p[0][8x16] & 4.19\\
                   sad\_sp\_x4[32x16] & 3.52 &              mean\_diff\_res[0][8x32] & 3.87 &                 subtract\_res[16x32] & 4.19\\
                  sub\_weight[16x32] & 3.53 &                     sad\_sp\_x4[64x8] & 3.88 &             mean\_diff\_res[1][64x16] & 4.2\\
              mean\_diff\_p[1][16x16] & 3.55 &                    block\_zero[64x8] & 3.9 &                mean\_diff\_p[0][8x32] & 4.22\\
                  sub\_weight[16x64] & 3.55 &              mean\_diff\_res[1][8x32] & 3.92 &                  copy\_bipred[16x64] & 4.23\\
                    sad\_sp\_x4[16x8] & 3.56 &              mean\_diff\_res[0][32x8] & 3.92 &                    sad\_sp\_x3[16x16] & 4.25\\
              mean\_diff\_p[0][64x16] & 3.57 &                    rdo\_clipmul[8x4] & 3.93 &                   block\_zero[64x64] & 4.29\\
\hline
\end{tabular}
\end{table}
\newpage
\begin{table}
\fontsize{8pt}{10pt}\selectfont
\begin{tabular}{| p{3.3cm}| p{0.7cm}|  p{3.3cm}|  p{0.7cm}|  p{3.3cm}|  p{0.7cm}| }
\hline
Primitive & IPC \newline Gain & Primitive & IPC \newline Gain & Primitive & IPC \newline Gain\\
\hline
           luma\_inv\_quant\_12b[8x32] & 6.95 &           luma\_inv\_quant\_12b[32x32] & 7.8 &            luma\_inv\_quant\_12b[8x64] & 8.4\\
           luma\_inv\_quant\_10b[8x32] & 6.97 &                    add\_avg\_12[16x2] & 7.84 &                       copy\_block[5] & 8.43\\
                       ssd\_sp[16x4] & 7 &                    luma\_vss\_10[8x4] & 7.84 &                    luma\_vss\_10[4x8] & 8.49\\
                   luma\_vps\_12[4x8] & 7 &                   luma\_vps\_10[4x16] & 7.87 &           luma\_inv\_quant\_12b[16x16] & 8.49\\
                       ssd\_pp[32x4] & 7.02 &                    luma\_vsp\_10[4x4] & 7.88 &           luma\_inv\_quant\_10b[32x16] & 8.51\\
                   luma\_vpp\_10[4x4] & 7.15 &           luma\_inv\_quant\_10b[64x64] & 7.93 &                 chroma\_vss\_12[16x2] & 8.64\\
                   luma\_vpp\_12[4x4] & 7.17 &                    add\_avg\_10[16x2] & 7.95 &           luma\_inv\_quant\_12b[16x64] & 8.65\\
\hline
\end{tabular}
\end{table}
\newpage
\begin{table}
\fontsize{8pt}{10pt}\selectfont
\begin{tabular}{| p{3.3cm}| p{0.7cm}|  p{3.3cm}|  p{0.7cm}|  p{3.3cm}|  p{0.7cm}| }
\hline
Primitive & IPC \newline Gain & Primitive & IPC \newline Gain & Primitive & IPC \newline Gain\\
\hline
              rdo\_num\_nonzero[16x8] & 10.94 &                chroma\_vps\_10[32x32] & 11.58 &                        ssd\_ss[32x4] & 12.13\\
                     satd\_sp[32x16] & 11 &                 chroma\_vpp\_12[16x2] & 11.59 &                  luma\_vsp\_10[64x32] & 12.15\\
                   chroma\_hps\_10[3] & 11 &                      luma\_hps\_10[4] & 11.6 &                    luma\_vpp\_10[8x8] & 12.17\\
                 luma\_vsp\_10[16x64] & 11 &                chroma\_vsp\_10[32x16] & 11.63 &                  add\_clip\_10[16x16] & 12.21\\
               ang\_fast\_copy[64x16] & 11.01 &                  luma\_vsp\_12[16x64] & 11.64 &                chroma\_vss\_12[16x16] & 12.24\\
                       ssd\_sp[16x8] & 11.04 &                      satd\_pp[32x64] & 11.66 &                chroma\_vpp\_12[32x64] & 12.26\\
                     satd\_sp[16x32] & 11.04 &                  luma\_vps\_10[32x32] & 11.68 &                   luma\_vss\_12[8x32] & 12.27\\
\hline
\end{tabular}
\end{table}
\newpage
\begin{table}
\fontsize{8pt}{10pt}\selectfont
\begin{tabular}{| p{3.3cm}| p{0.7cm}|  p{3.3cm}|  p{0.7cm}|  p{3.3cm}|  p{0.7cm}| }
\hline
Primitive & IPC \newline Gain & Primitive & IPC \newline Gain & Primitive & IPC \newline Gain\\
\hline
                 luma\_vsp\_12[64x32] & 14.35 &                   luma\_vpp\_10[32x8] & 14.98 &                  add\_clip\_10[64x64] & 15.52\\
                       ssd\_ss[32x8] & 14.41 &                      luma\_hps\_10[3] & 15.01 &                   add\_clip\_12[32x8] & 15.54\\
                 luma\_vpp\_10[16x16] & 14.43 &                      luma\_hpp\_10[4] & 15.01 &                       ssd\_ss[32x32] & 15.56\\
                      ssd\_sp[32x16] & 14.44 &                 chroma\_vps\_10[32x8] & 15.02 &                chroma\_vps\_10[16x16] & 15.56\\
                chroma\_vps\_10[16x8] & 14.44 &                   add\_clip\_10[64x8] & 15.07 &                chroma\_vss\_10[32x16] & 15.58\\
                chroma\_vpp\_10[32x4] & 14.45 &                chroma\_vpp\_12[32x32] & 15.09 &                chroma\_vss\_12[32x64] & 15.58\\
                  luma\_vpp\_12[8x32] & 14.49 &                  luma\_vsp\_12[64x64] & 15.09 &                chroma\_vpp\_12[32x16] & 15.59\\
\hline
\end{tabular}
\end{table}